\documentclass[12pt]{article}
\usepackage{cite}
\usepackage[T1]{fontenc}
\usepackage[utf8]{inputenc}

\usepackage[english]{babel}

\usepackage{graphicx}
\usepackage[dvipsnames]{xcolor}
\usepackage{caption}
\usepackage{multirow}
\usepackage{subfig}
\usepackage{float}

\usepackage{amsmath}
\DeclareMathOperator{\Tr}{Tr}
\usepackage{amsfonts}
\usepackage{amssymb}
\usepackage{amstext}
\usepackage{slashed}

\usepackage{multirow}
\usepackage{booktabs}

\usepackage[bookmarks, breaklinks, colorlinks,urlcolor=black, citecolor=red, 
linkcolor=blue]{hyperref}
\usepackage{placeins}
\usepackage{verbatim}

\usepackage{float} % For table position

\begin{document}
	
	\renewcommand*{\thefootnote}{\fnsymbol{footnote}}
	
	%%%%%%%%%%%%%%%%%%%%%%%%%%%%%%%%%%%%%%%%%%%%%%%%%%%%%%%
\mbox{}\hfill{IFIC/19-43}
\vskip 2cm	
	
	\begin{center}
		
{\Large\bf 
LHC bounds on coloured scalars} 
\\[10mm]
{V\'ictor Miralles\footnote{Email: Victor.Miralles@ific.uv.es}
and Antonio Pich}\footnote{Email: Antonio.Pich@ific.uv.es}

{\small\em IFIC, Universitat de Val\`encia-CSIC, Apt. Correus 22085, E-46071 Val\`encia, Spain} \\[3mm]

	\end{center}
%%%%%%%%%%%%%%%%%%%%%%%%%%%%%%%%%%%%%%%%%%%%%%%%%%%%%%%%%%%%%%

\begin{abstract}

We analyze the constraints on coloured scalar bosons imposed by the current LHC data at $\sqrt{s}=13$~TeV. Specifically, we consider an additional electroweak doublet of colour-octet scalars, satisfying the principle of minimal flavour violation in order to fulfill the stringent experimental limits on flavour-changing neutral currents. We demonstrate that coloured scalars with masses below 800~GeV are already excluded, provided they are not fermiophobic.

\end{abstract}

%\clearpage
%
%\begin{spacing}{0}
%\tableofcontents
%\end{spacing}

\section{Introduction}

The discovery of the Higgs boson by the ATLAS and CMS Collaborations \cite{ATLASHiggs,CMSHiggs} can be understood not just like the confirmation of the mechanism for generating the masses of the Standard Model (SM) particles but like the discovery of the first element of a possible more extended scalar sector. Although the current data are compatible with the SM, there remain many unanswered fundamental questions which motivate the existence of new physics (NP). Furthermore, with the available data, there is still plenty of room for extensions of the scalar sector of the SM at the TeV scale. 

Electroweak models with extended scalar sectors usually introduce two potentially worrisome problems: dangerous contributions to the mass ratio $\rho\equiv M_W^2/(M_Z^2\cos^2{\theta_W})$ and unsuppressed flavour-changing neutral-current (FCNC) transitions. The phenomenological requirement that $\rho- 1$ must be very small, $\le \mathcal{O}(10^{-3})$, selects $SU(2)_L$ singlets and doublets as the preferred scalar candidates, while unwanted FCNCs are usually avoided introducing some {\it ad hoc} discrete symmetry that 
forces each type of SM right-handed fermion to couple only to a single scalar doublet. This guarantees that all Yukawa matrices are diagonal in the mass basis and keeps the resulting flavour structure stable under quantum corrections (natural flavour conservation) \cite{gw76}. However, there are more generic possibilities in order to suppress FCNCs.

The principle of minimal flavour violation (MFV)  \cite{C1987,DAmbrosio:2002vsn} constitutes a much weaker (and general) assumption that also leads to a very effective suppression of FCNCs. It is based on the hypothesis that all Yukawa matrices are proportional to the same flavour structures that break the $SU(3)_{Q_L}\otimes SU(3)_{u_R}\otimes SU(3)_{d_R}$ symmetry. In multi-Higgs doublet models this leads to the flavour alignment of all scalar Yukawa couplings to a given right-handed fermion type \cite{Pich:2009sp,Pich:2010ic,Penuelas:2017ikk}. FCNCs are then absent at tree level, and all flavour-changing phenomena are controlled by the charged-current Cabibbo-Kobayashi-Maskawa (CKM) matrix \cite{Cabibbo:1963yz,Kobayashi:1973fv}. 

Manohar and Wise realized that the principle of MFV 
can only be satisfied by those scalar representations transforming under $SU(3)_C\otimes SU(2)_L\otimes U(1)_Y$ like ({\bf 1,2})$_{1/2}$ or ({\bf 8,2})$_{1/2}$ \cite{mw06}. Therefore models with additional scalar doublets that are either colour singlet or colour octet
become very interesting candidates for possible extensions of the scalar sector of the SM. The phenomenological implications of having additional colour-singlet scalar doublets have been extensively studied \cite{Jung:2010ik,Jung:2010ab,Ferreira:2010xe,Braeuninger:2010td,Cree:2011uy,Serodio:2011hg,Bijnens:2011gd,Jung:2012vu,Celis:2012dk,Altmannshofer:2012ar,Bai:2012ex,Akeroyd:2012yg,Celis:2013rcs,Jung:2013hka,Celis:2013ixa,Duarte:2013zfa,Wang:2013sha,Lopez-Val:2013yba,Li:2014fea,Ilisie:2014hea,Mileo:2014pda,Ilisie:2015tra,Abbas:2015cua,Chang:2015rva,Han:2015yys,Enomoto:2015wbn,Celis:2016azn,Cherchiglia:2016eui,Hu:2016gpe,Ayala:2016djv,Penuelas:2017ikk,Hu:2017qxj,Cho:2017jym,Gori:2017qwg,Chowdhury:2017aav,Cherchiglia:2017uwv,Grzadkowski:2018ohf}. In this work we focus on the other possibility, scalar extensions in which we add a colour-octet electroweak doublet satisfying the principle of MFV, the so-called Manohar and Wise (MW) model \cite{mw06}. This model is also motivated by the fact that some $SU(5)$ and $SO(10)$ unification theories predict colour-octet electroweak doublets with masses around the electroweak scale \cite{Georgi:1979df,Pati:1983zp,Dorsner:2006dj,Perez:2013osa,Perez:2016qbo,Bertolini:2013vta}.

The MW model has also been widely studied in the literature,
and its parameter space has been constrained with theoretical considerations, such as unitarity and vacuum stability \cite{hhyv13,Cheng:2018mkc}, and phenomenological analyses.
The presence of colour-octet scalars affects the SM Higgs production and its decay to diphotons \cite{He:2011ti,Dobrescu:2011aa,Bai:2011aa,Cao:2013wqa,Cheng:2016tlc}, 
electroweak precision observables like the oblique parameters \cite{mw06,btz09} or $R_b$ \cite{gw07,Degrassi:2010ne}, 
flavour observables such as neutral-meson mixing or the decay $B_s\to\ell^+\,\ell^-$ \cite{clyz15}, and the anomalous electric and magnetic dipole moments of the quarks \cite{mv16}.

Being coloured particles, the MW scalars could be massively produced at the LHC,  
provided they are light enough to be kinematically accessible. However, most of the works analysing the direct production of these scalars have been performed 
before the huge release of data on heavy-particle searches at the LHC \cite{Gerbush:2007fe,Burgess:2009wm,Arnold:2011ra,Kribs:2012kz,Hayreter:2017wra,Hayreter:2018ybt,Darme:2018dvz}.
The most complete analysis on the direct production of these scalars was done in Ref.~\cite{Hayreter:2017wra}, where the single and pair production of neutral scalars was studied.
Slightly stronger limits were obtained in Ref.~\cite{Darme:2018dvz} for the particular case of a neutral coloured pseudoscalar, assuming that it decays exclusively into top quarks (top-philic limit).
The recent release of high-luminosity experimental searches for massive resonances decaying into heavy quarks \cite{Aaboud:2018mjh, Aaboud:2018xpj, Aaboud:2018cwk,Sirunyan:2019wxt} 
makes it now possible to improve those bounds. In this work, we focus on the associated production of neutral and charged scalars with heavy quarks that was not considered in previous analyses. With these searches we are able to cover some regions of the parameter space that have not been studied before.
Furthermore, we also update the experimental data on the single production of neutral scalars decaying to top quarks, at $\sqrt{s}=13$ TeV. However, the limits obtained from this channel are still not better than those obtained for $\sqrt{s}=8$ TeV in Ref. \cite{Hayreter:2017wra}. 

Note that in order to compare our theoretical predictions with the experimental data we make use of specific analyses performed by the ATLAS and CMS Collaborations. In the case of the single production, we compare our predictions with the experimental limits found for $Z'$ particles and Kaluza-Klein gluons and gravitons \cite{Aaboud:2018mjh}. For the associated production, we exploit the limits obtained for the type-II two-Higgs-doublet model (2HDM) in Refs. \cite{Aaboud:2018xpj, Aaboud:2018cwk, Sirunyan:2019wxt}. We have considered that the kinematics of our production channels and the experimental analyses we compare with are very similar, which is a reasonable assumption.

Although in this work we have focused in the MW model, it is worth to mention that there are other possibilities to suppress the NP contributions to FCNCs, like the next-to-minimal-flavour-violation scenario where the NP dominantly couples with the third-generation quarks and is quasialigned with the Yukawa matrices \cite{Agashe:2005hk}. Phenomenological signatures of coloured-octet scalars in this type of models have been studied in Refs.~\cite{Chivukula:2013kw,Chivukula:2013hga}.

We first provide in Section~\ref{sec:MW} a very brief description of the MW model and define the relevant parameters. Our phenomenological analyses are detailed in Section~\ref{sec:Phenomenology}, where we analyze the single production of neutral scalars and the associated production of neutral and charged scalars with top quarks. We finally summarize in Section~\ref{sec:Summary}.

\section{The MW Model}
\label{sec:MW}

As we mentioned in the Introduction, the MW model adds a new scalar field to the SM with the $SU(3)_C\otimes SU(2)_L\otimes U(1)_Y$ quantum numbers ({\bf 8,2})$_{1/2}$. Therefore the scalar sector will be formed by the usual Higgs doublet  $\phi = (\phi^+ , \phi^0)^T $ plus an $SU(3)_C$-octet field $S^A = (S^{A,+} , S^{A,0})^T$. Since they are colourful particles, the new scalars cannot acquire a vacuum expectation value (vev), neither can they mix with the SM Higgs doublet.

The most general potential that can be build with this scalar sector takes the form:
\begin{align}
\nonumber V&=\frac{\lambda}{16}\, (2\,\phi^{\dagger i}\phi_i-v^2)^2 + 2m_S^2\,\Tr ( S^{\dagger i}S_i) +\lambda_1\,\phi^{\dagger i}\phi_i \Tr (S^{\dagger j}S_j) +\lambda_2\,\phi^{\dagger i}\phi_j \Tr (S^{\dagger j}S_i) 
\\ \nonumber &
+ \left[\lambda_3\,\phi^{\dagger i}\phi^{\dagger j} \Tr (S_iS_j)
%%% \\\nonumber&
+\lambda_4\,\phi^{\dagger i} \Tr (S^{\dagger j}S_jS_i) +\lambda_5\,\phi^{\dagger i} \Tr (S^{\dagger j}S_iS_j) +\text{h. c.}\right]
\\[2pt] \nonumber &
+\lambda_6\, \Tr (S^{\dagger i}S_iS^{\dagger j}S_j)
+\lambda_7 \,\Tr (S^{\dagger i}S_jS^{\dagger j}S_i)
%%%\\&
+\lambda_8\, \Tr (S^{\dagger i}S_i) \Tr (S^{\dagger j}S_j) 
\\[2pt] &
+ \lambda_9\, \Tr (S^{\dagger i}S_j) \Tr (S^{\dagger j}S_i) +\lambda_{10}\, \Tr (S_iS_j) \Tr (S^{\dagger i}S^{\dagger j}) +\lambda_{11}\, \Tr (S_iS_j S^{\dagger j}S^{\dagger i})\, ,
\label{eq:octetpot}
\end{align}
where $i$ and $j$ are $SU(2)_L$ indices, the traces are in colour space and we have used the notation $S=S^AT^A$, with $T^A$ the generators of the $SU(3)_C$ group. All potential parameters are real except $\lambda_3$, $\lambda_4$ and $\lambda_5$, but we can choose $\lambda_3$ to be real performing a global phase rotation of the $S$ multiplet field.

The vev of the SM Higgs doublet, $\langle\phi^0\rangle =v/\sqrt{2}$, generates a mass splitting among the physical coloured scalars,

\begin{equation}
m_{S^\pm}^2=m_S^2+\lambda_1\,\frac{v^2}{4}\, ,
\hspace{1.5cm} 
m_{S^{0}_{R,I}}^2=m_S^2+(\lambda_1+\lambda_2\pm2\lambda_3)\, \frac{v^2}{4}\, ,
\end{equation}
where $m_{S^\pm}$ is the charged-scalar mass, while $m_{S^{0}_{R}}$ and $m_{S^{0}_{I}}$ are the masses of the CP-even and CP-odd neutral scalars, respectively.

The interaction of the octet scalars with the gauge bosons is generated by the kinetic term
\begin{equation}\label{eq:kin}
\mathcal{L}_{K}\, =\, 2\Tr[(D_\mu S)^\dagger D^\mu S]\, ,
\end{equation}
through the covariant derivative,
\begin{equation}
D_\mu S\, =\, \partial_\mu S+ i g_s\, [G_\mu,S]+
i g\,\frac{\sigma^i}{2} W_\mu^i S
+\frac{i}{2} g'B_\mu S \, ,
\end{equation}
with $G_\mu = G_\mu^A T^A$ the octet gluon field.
The factor of 2 in Eq.~(\ref{eq:kin}) generates the correct canonical normalisation for the fields.

The last remaining piece is the Yukawa interaction 
of the colour-octet scalar multiplet. The MFV assumption implies that the Yukawa flavour matrices of the $S$ field are proportional to the SM ones,
\begin{equation}
\mathcal{L}_{Y}=-\sum^3_{i,j=1}\Big[\eta_D\, Y^d_{ij}\,\overline{Q}_{L_i}S d_{R_j}+ \eta_U\, Y^u_{ij}\,\overline{Q}_{L_i}\widetilde{S}u_{R_j}+\text{H.c.}\Big] \, ,
\label{eq:yuk}
\end{equation}
where $\eta_D$ and $\eta_U$ are, in general, complex parameters. The scalar-fermion interactions are then proportional to the corresponding fermion masses since $Y^f = \sqrt{2}\, M_f/v$.

The $S^0_{R,I} G^2$ interaction plays an important role in the production and decay of the coloured neutral scalars. The dimension-4 Lagrangian does not contain any direct coupling of the neutral octet scalars to two gluon fields. However, this coupling gets generated by quantum effects through scalar and fermion loops. The corresponding vertex can be represented by the dimension-6 gauge-invariant effective Lagrangian,
\begin{equation}
\mathcal{L}_{SGG}= F_R \; G^A_{\mu\nu}G^{B\mu\nu}d^{ABC}S^{0^C}_R+F_I \; \widetilde{G}^A_{\mu \nu}G^{B\mu\nu}d^{ABC}S^{0^C}_I \, ,
\label{eq:eft}
\end{equation}
where $G^A_{\mu\nu}$ is the gluon strength tensor and $\widetilde{G}^A_{\mu \nu}=\frac{1}{2}\epsilon^{\mu\nu\alpha\beta}G^A_{\alpha\beta}$.
The Wilson coefficients $F_R$ and $F_I$ are easily obtained at the one-loop level.
This was first calculated in  Ref.~\cite{gw07}, neglecting the mass splitting between the scalars. If the splitting is not neglected, these coefficients are given by
 \begin{align}
\nonumber& F_R\, =\, (\sqrt{2} G_F)^{1/2}\;\frac{\alpha_s}{8 \pi} \;\Bigg[\eta_U\, I_q \bigg(\frac{m_t^2}{m_{S_R}^2}\bigg)+\eta_D\, I_q \bigg(\frac{m_b^2}{m_{S_R}^2}\bigg)
\\\nonumber& \hskip 3.9cm\mbox{}
-\frac{9}{4}\frac{v^2}{m_{S_R}^2}\frac{\lambda_4+\lambda_5}{2}\,\left\{ I_s (1)+\frac{1}{3}\,\bigg[ I_s \bigg(\frac{m_{S_I}^2}{m_{S_R}^2}\bigg)+2 I_s \bigg(\frac{m_{S^\pm}^2}{m_{S_R}^2}\bigg)\bigg]\right\}\Bigg],
\\&
F_I\, =\, (\sqrt{2} G_F)^{1/2}\;\frac{\alpha_s}{16 \pi} \;\Bigg[ \eta_U\,\frac{m_t^2}{m_{S_I}^2}\, \mathcal{F}\bigg(\frac{m_t^2}{m_{S_I}^2}\bigg)+ \eta_D\,\frac{m_b^2}{m_{S_I}^2}\, \mathcal{F}\bigg(\frac{m_b^2}{m_{S_I}^2}\bigg)\Bigg] \, ,
\label{eq:wilsoncoef}
\end{align}
in the CP-conserving limit, i.e., considering all parameters to be real. In these expressions we have made use of the functions, 
\begin{equation}
I_q(z)\, =\, z\,\left[ 2 + (4z-1) \,\mathcal{F}(z)\right]\, ,
\qquad \qquad
I_s(z)\, =\, -z\,\left[ 1+2z \,\mathcal{F}(z)\right]\, ,
\end{equation}
where
\begin{equation}
\mathcal{F}(z)\, =\,\left\{ 
\begin{array}{ccc}
  \frac12 \left[\log{\left(\frac{1+\sqrt{1-4z}}{1-\sqrt{1-4z}}\right)}-i\pi\right]^2 , 
  &\hskip 1cm & z<1/4
 \\[12pt]
-2\,\arcsin^2(1/\sqrt{4z})\, , && z>1/4
  \end{array} \right.\, .
\end{equation}

In order to reduce the total number of parameters, we will work in the CP-conserving limit. Furthermore, the couplings $m^2_S$, $\lambda_{1}$,  $\lambda_{2}$ and $\lambda_{3}$ are only needed for the determination of the scalar masses; since we only have three different coloured scalars, we can then remove 1 degree of freedom. Moreover, the parameters $\lambda_4$ and $\lambda_5$ are just relevant for the decay of the CP-even neutral scalar to gluons, which depends on their sum $\lambda_4 + \lambda_5$, allowing us to remove another degree of freedom. Finally, the four-point interactions of the coloured scalars will be irrelevant for this analysis, so we can take $\lambda_{6-11}=0$ for simplicity.

With this considerations, we end up with only 6 degrees of freedom,
\begin{align}
\nonumber& m^2_{S^\pm}\, ,\qquad  \lambda_{4,5}=\frac{\lambda_{4}+\lambda_{5}}{2}\, , \qquad \eta_U\, , \qquad \eta_D\, ,\\\nonumber&\Delta m^2_{S_R}= m^2_{S^0_R} - m^2_{S^\pm}=\frac{v^2}{4}(\lambda_{2}+2\,\lambda_{3})\, ,\\& \Delta m^2_{S_I}= m^2_{S^0_I} - m^2_{S^\pm}=\frac{v^2}{4}(\lambda_{2}-2\,\lambda_{3})\, .
\end{align}
These parameters must satisfy some theoretical requirements and phenomenological constraints. Perturbative unitarity enforces the modulus of $\lambda_{4,5}$ to be smaller than 13 \cite{hhyv13}. Flavour observables~ \cite{clyz15} and $R_b$~\cite{gw07} put an upper bound on the up-type Yukawa, $|\eta_U|<2$, for a charged-scalar mass below 1~TeV. The analogous limits on $|\eta_D|$ are currently very weak, allowing it to go beyond 100. Finally, the mass splittings are constrained by the oblique parameters S, T and U. For light scalar masses ($\le 350$~GeV), $\Delta m^2_{S_I}$ ($\Delta m^2_{S_R}$) is constrained to be smaller than 60 GeV if $\lambda_2$ and $\lambda_3$ have the same (opposite) sign~\cite{Burgess:2009wm}. The oblique constraint becomes weaker for larger masses.

\section{Collider Phenomenology}
\label{sec:Phenomenology}

In our phenomenological analysis, we have considered the processes
$p\, p \rightarrow S^0_{R,I} \rightarrow t \,\bar{t}$,  $p\, p \rightarrow S^0_{R,I}\, t \,\bar{t}\rightarrow t \,\bar{t}\;  t \,\bar{t}  $ and $p\, p \rightarrow S^+\, \bar t \, b\rightarrow t \,\bar{b}\;  \bar t \, b$. In order to 
constrain the model, we will focus on those regions of the parameter space in which the coloured scalar particles decay mainly to quarks, i.e., where decay modes into another coloured scalar and a weak boson such as $S^0_{R,I} \rightarrow S^+ W^-$, $S^+ \rightarrow S_{R,I}^0 W^+$ and $S_{R,I}^0 \rightarrow S_{I,R}^0 Z^0$ are suppressed. 
Of course, these channels are only possible for the heavier scalars decaying to the lighter ones because the opposite possibility is kinematically forbidden. Therefore, when we consider the production of every scalar, we select $\lambda_2$ and $\lambda_3$ in such a way that this scalar is the lightest one, avoiding then the unwanted decay modes. Hence, the study of the CP-even neutral scalars, the one of the CP-odd neutral scalars and the one of the charged scalars analyze different regions of $\lambda_2$ and $\lambda_3$, but once we combine the results coming from the three searches, we are able to cover all possible values of these parameters. 

We have generated collision events with the program MG5$\_$aMC@NLO \cite{a14mad}, using first  {\sc feynrules } \cite{c08,d11} to produce the universal Feynrules Output needed to run our model. All calculations in MG5aMC@NLO were performed at tree level. 
In the event generation we have used the particle distribution functions (PDF) set NNPDF23\_nlo\_ as\_0119 of the package {\sc lhapdf}\_6.1.6 \cite{b14}. The centre-of-mass energy has been fixed to $\sqrt{s}=13$~TeV, while the experimental data had an integrated luminosity of 36.1 fb$^{-1}$ for the ATLAS data  \cite{Aaboud:2018mjh, Aaboud:2018xpj, Aaboud:2018cwk} and 137 fb$^{-1}$ for the CMS data \cite{Sirunyan:2019wxt}. Furthermore, we have estimated the theoretical uncertainties varying the values of the renormalisation and factorisation scales by a factor between $1/2$ and $2$.

\subsection{Single Production of Neutral Scalars}

The neutral coloured scalars, apart from decaying into top quarks, can also decay to gluons and to bottom quarks. The decay amplitude to gluons depends on the parameters $\eta_U$ and $\eta_D$ for the CP-odd scalars, and also on $\lambda_{4,5}$ for the CP-even scalars; the decay to $t\, \bar{t}$ is proportional to $\eta_U$ and the decay to $b\,\bar{b}$ is proportional to $\eta_D$.  Furthermore, as mentioned before, we select the values of $\lambda_2$ and $\lambda_3$ in such a way that the decay into another coloured scalar is kinematically forbidden, i.e., $\lambda_2+2\lambda_3<0$ and $\lambda_3<0$ for the CP-even scalar, and $\lambda_2-2\lambda_3<0$ and $\lambda_3>0$ for the CP-odd scalar. 

In order to maximize the production cross section of a coloured neutral CP-even scalar through gluon fusion, we have chosen $\lambda_{4,5}$ with the opposite sign to $\eta_U$ and $\eta_D$, so that all contributions to $F_R$ in Eq.~\eqref{eq:wilsoncoef} interfere constructively. Notice that this Wilson coefficient also governs the $S^0_R$ decay into two gluons. Hence, both the single production cross section and the decay amplitude to gluons are higher in this case compared with the alternative sign choice for $\lambda_{4,5}$. However, as shown in Ref. \cite{Hayreter:2017wra}, the scalar contribution is much smaller than the top one, unless $|\lambda_{4,5}|$ is very large. Therefore, one gets finally similar results with both signs. 
The resulting limits are just slightly better with our choice of sign due to the increase of the cross section; the corresponding increase of the decay width into gluons does not worsen them because the branching ratio of the $S^0_R$ decay into quarks remains always very close to one, for the values of $\eta_U$ considered here. We have varied $|\lambda_{4,5}|$ from 0 to $10$, which is almost at the limit of its perturbative unitarity region ($|\lambda_{4,5}| \lesssim 13$)  \cite{hhyv13}. As it has been previously commented, the parameter $\eta_U$ is strongly constrained by flavour observables and, for masses of the coloured scalars smaller than 1 TeV, its absolute value cannot be higher than 2 \cite{gw07,clyz15}, so we go at most up to this value of $|\eta_U|$.

For the single production we have taken the experimental data from the ATLAS search for $Z'$ bosons and Kaluza-Klein gravitons and gluons, in Ref.~\cite{Aaboud:2018mjh}.
The interference with the SM production amplitudes  has not been considered in those searches, although it could have some effects on the signal shape for scalars and pseudoscalars decaying to top quarks \cite{Aaboud:2017hnm}. Therefore, some care has to be taken with the limits obtained from these channels, until more direct searches for coloured scalars and pseudoscalars are released.
Note also that for these searches we have not considered any QCD corrections. In the SM these corrections enhance the single production of the Higgs boson at the LHC by a factor around 1.5 \cite{Spira:1995rr}, and we could expect a similar or even larger contribution for our coloured scalars. However, there are additional Feynman diagrams contributing to the production of the coloured scalars, which have not been calculated yet. Since the missing QCD corrections are expected to increase the production cross section, our limits are then quite conservative.

%%%%%%%%%%%%%%%%%%%%%%%%%%% Figure %%%%%%%%%%%%%%%%%%%%%%%%%%%%%%%
 \begin{figure}[tb]
\centering
\subfloat[]{
\includegraphics[scale=0.5]{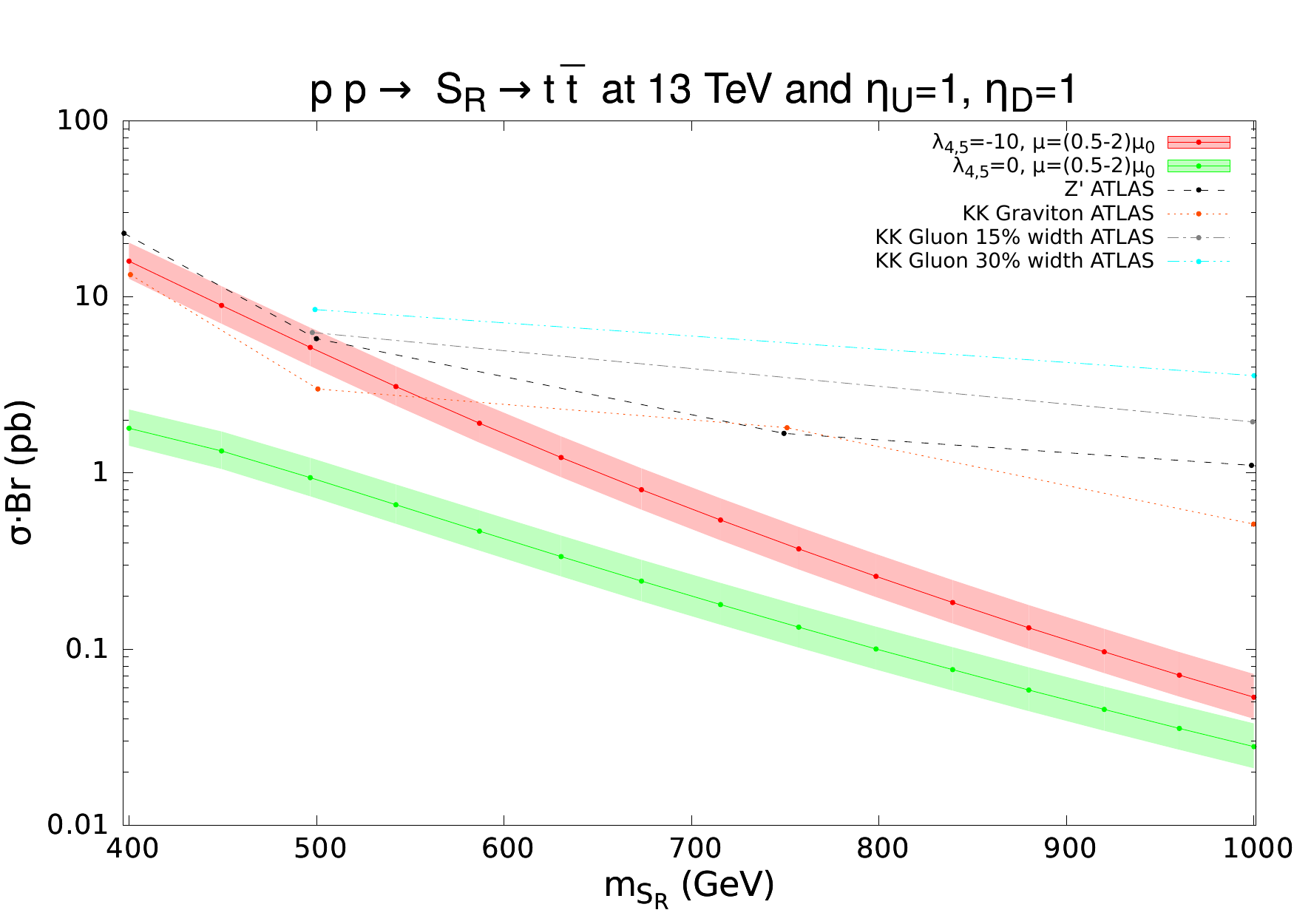}
}
\subfloat[]{
\includegraphics[scale=0.5]{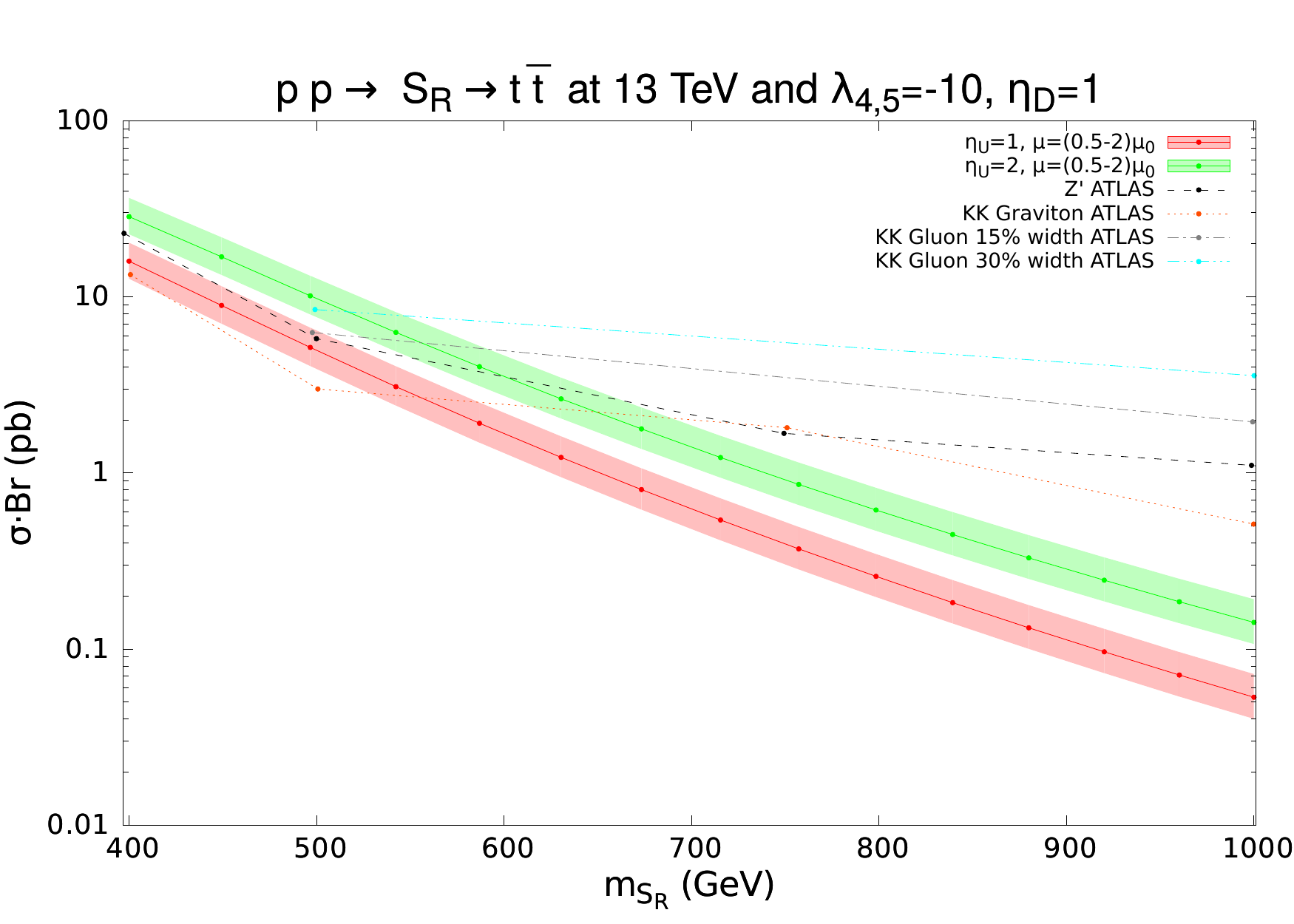}
}
\caption{Cross section times branching ratio for the single production of $S^0_R$ and its decay to $t\bar t$. In (a) we set $\eta_U$ to 1 and vary $\lambda_{4,5}$ from $-10$ to 0 and in (b) $\lambda_{4,5}= -10$ and $\eta_U$ is varied from 1 to 2. The nonsolid lines correspond to the experimental limits from Ref.~\cite{Aaboud:2018mjh}.}
\label{fig:1S2t} 
\end{figure}
%%%%%%%%%%%%%%%%%%%%%%%%%%%%%%%%%%%%%%%%%%%%%%%%%%%%%%%%%%%%%%%%%%%%%%%

Figure~\ref{fig:1S2t} compares the (95\% C.L.) experimental limits on the production of heavy particles that decay into top-quark pairs~\cite{Aaboud:2017hnm} with the calculated production cross section times branching ratio for the CP-even coloured scalar, as a function of the scalar mass. The model-dependence of the ATLAS exclusion limits can be appreciated from the broad range of bounds obtained for the different explicit models analyzed: $Z'$ bosons, Kaluza-Klein gluons and Kaluza-Klein gravitons decaying into $t\,\bar t$.
The production of CP-even scalars is dominated by the gluon-fusion mechanism, which depends on 
$\eta_U$, $\eta_D$ and $\lambda_{4,5}$. Therefore, this experimental constraint can be easily avoided, taking small-enough values for these parameters. When $|\eta_U|$ is of order one,
the branching ratio for the decay $S^0_R\to t\bar t$ is almost one, provided $m_{S^0_R} > 2 m_t$,
for all values of $\lambda_{4,5}$ within its perturbative unitarity region. Owing to the large mass splitting between the top and bottom quarks, the value of $\eta_D$ is also almost irrelevant for this branching ratio, so we have just taken $\eta_D=1$ as a representative value. 

It is evident from the figure that this channel does not provide strong constraints on the scalar mass. For $\eta_U=1$ and $\lambda_{4,5}=-10$, we can just infer that $m_{S^0_R}$ should be heavier than 500 GeV, comparing the predicted cross section with the ATLAS bound on the production of Kaluza-Klein gravitons. For smaller values of $\eta_U$ and/or $|\lambda_{4,5}|$, one does not obtain any useful constraints. 
Obviously, the missing QCD corrections could not change much the situation, which, furthermore, gets slightly worse with the opposite sign choice for $\lambda_{4,5}$.
In spite of the much higher statistics accumulated at 13 TeV, compared with the data sample analyzed in Ref.~\cite{Hayreter:2017wra}, the emerging limits are still not better than those extracted from the 8 TeV data and are not competitive with the ones obtained from the associated-production process, which are shown afterwards.

%%%%%%%%%%%%%%%%%%%%%%%%%%% Figure %%%%%%%%%%%%%%%%%%%%%%%%%%%%%%%%%%
 \begin{figure}[tb]
\centering
\subfloat[]{
\includegraphics[scale=0.5]{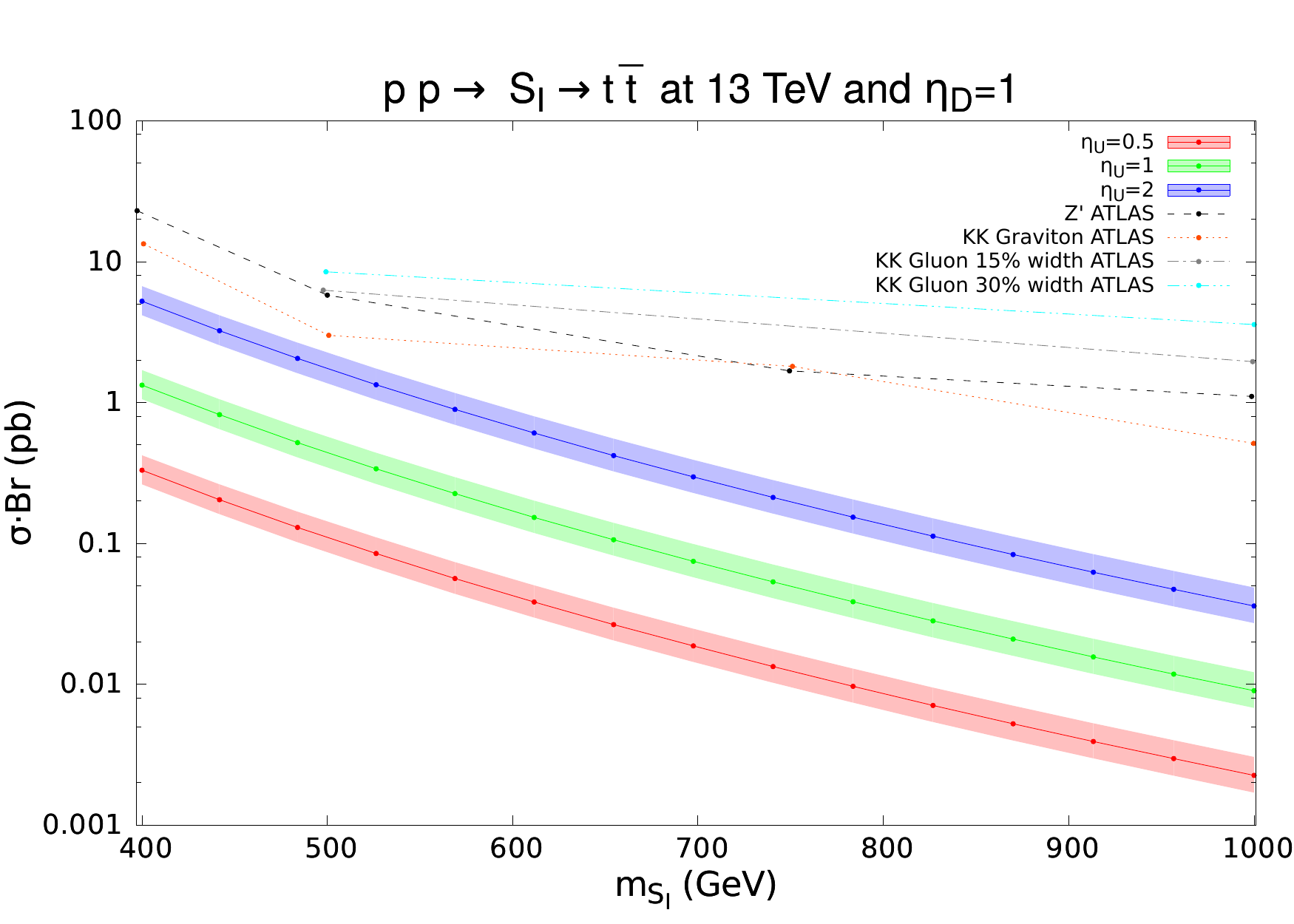}
}
\subfloat[]{
\includegraphics[scale=0.5]{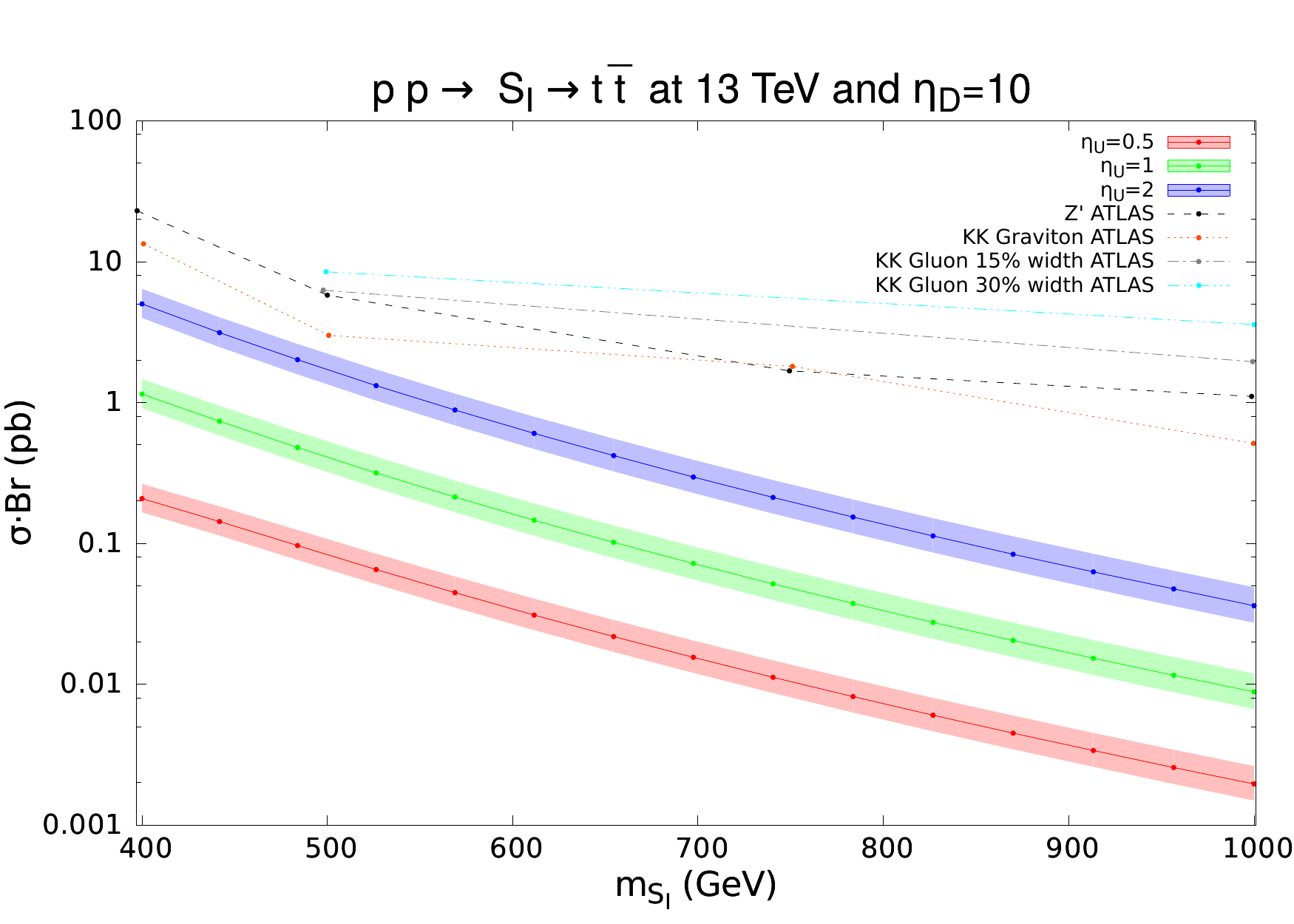}
}
\caption{Cross section times branching ratio for the single production of $S^0_I$ and its decay
to $t\bar t$. $\eta_U$ has been varied from 0.5 to 2, while  $\eta_D=1$ ($\eta_D=10$) in the left (right) panel. The nonsolid lines correspond to the experimental limits from Ref.~\cite{Aaboud:2018mjh}.}
\label{fig:1Si2t} 
\end{figure}
%%%%%%%%%%%%%%%%%%%%%%%%%%%%%%%%%%%%%%%%%%%%%%%%%%%%%%%%%%%%%%%%%%%%%%%%%

The analogous limits on the single production of the CP-odd scalar $S^0_I$ are shown in Fig.~\ref{fig:1Si2t}.
The behaviour is very similar to the CP-even case with $\lambda_{4,5}=0$, 
because the production of CP-odd scalars does not depend on $\lambda_{4,5}$. The production cross section depends only on $\eta_U$ and $\eta_D$, and it grows with the modulus of these parameters, although the dependence on $\eta_D$ is again extremely weak. The numerical differences between the left ($\eta_D=1$) and right ($\eta_D=10$) panels can hardly be seen in the figure.
For values of $|\eta_U|\le 2$, the predicted signal remains below the experimental limits in the whole range of $m_{S^0_I}$ analyzed. Therefore, this channel does not provide any constraint.

\subsection{Associated Production of Neutral Scalars and Top Quarks}

%%%%%%%%%%%%%%%%%%%%%%%%%%%%%% Figure %%%%%%%%%%%%%%%%%%%%%%%%%%%%%%%
\begin{figure}[tbh]
\centering
\subfloat[]{
\includegraphics[scale=0.28]{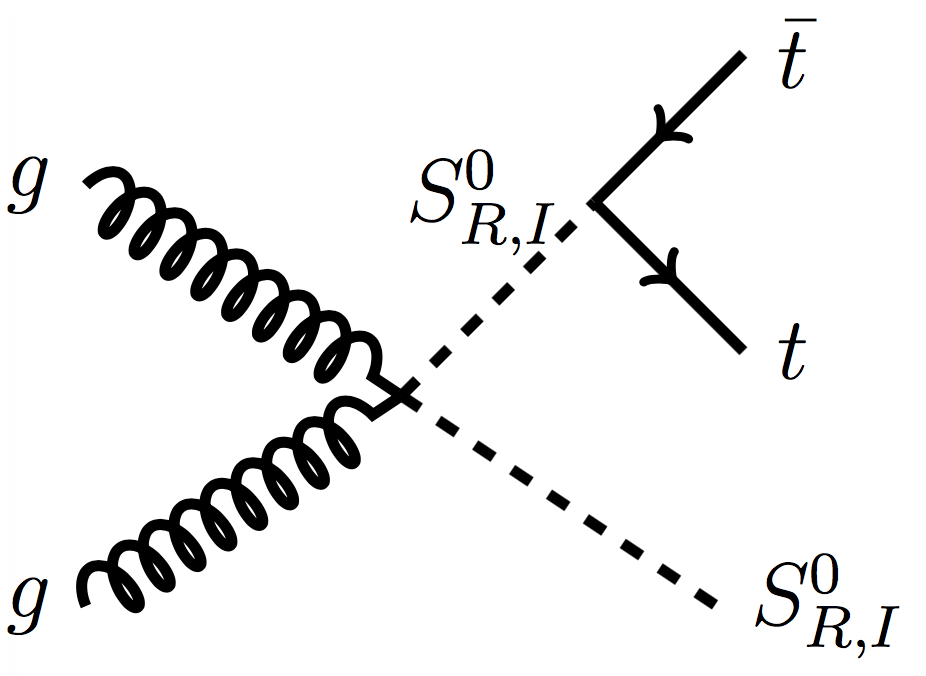}
}
\hspace{2.0 cm}
\subfloat[]{
\includegraphics[scale=0.28]{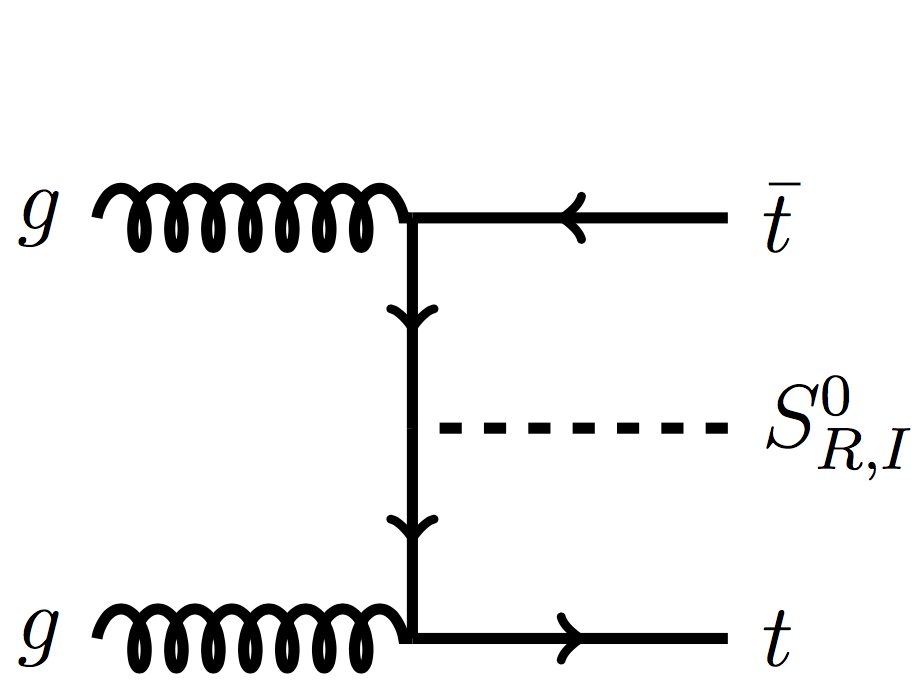}
}
\caption{Representative Feynman diagrams contributing to the associated production of neutral scalars with top quarks.}
\label{fig:Diagrams_Stt} 
\end{figure}

%%%%%%%%%%%%%%%%%%%%%%%%%%%%%% Figure %%%%%%%%%%%%%%%%%%%%%%%%%%%%%%
 \begin{figure}[t!]
\centering
\subfloat[]{
\includegraphics[scale=0.5]{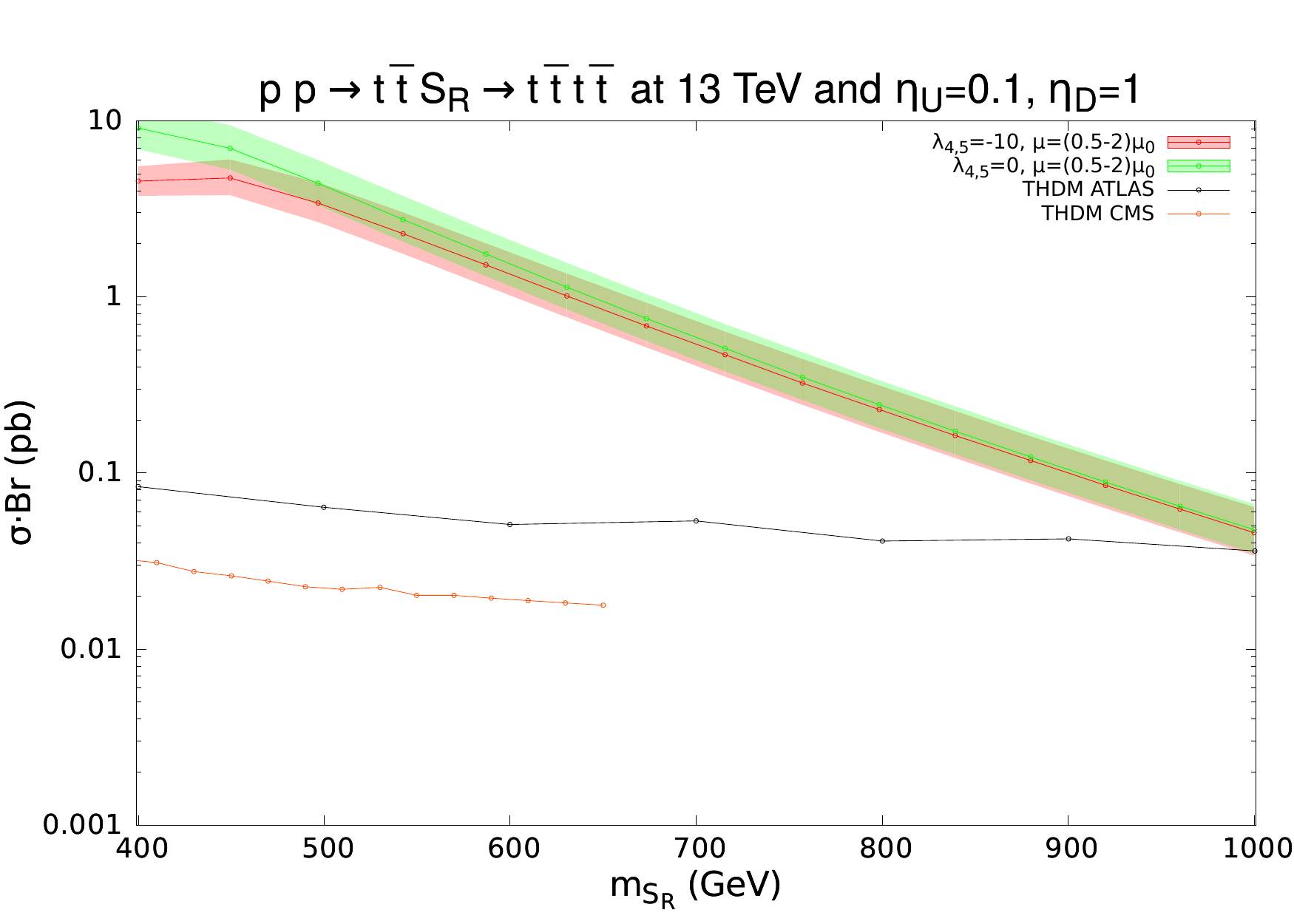}
}
\subfloat[]{
\includegraphics[scale=0.5]{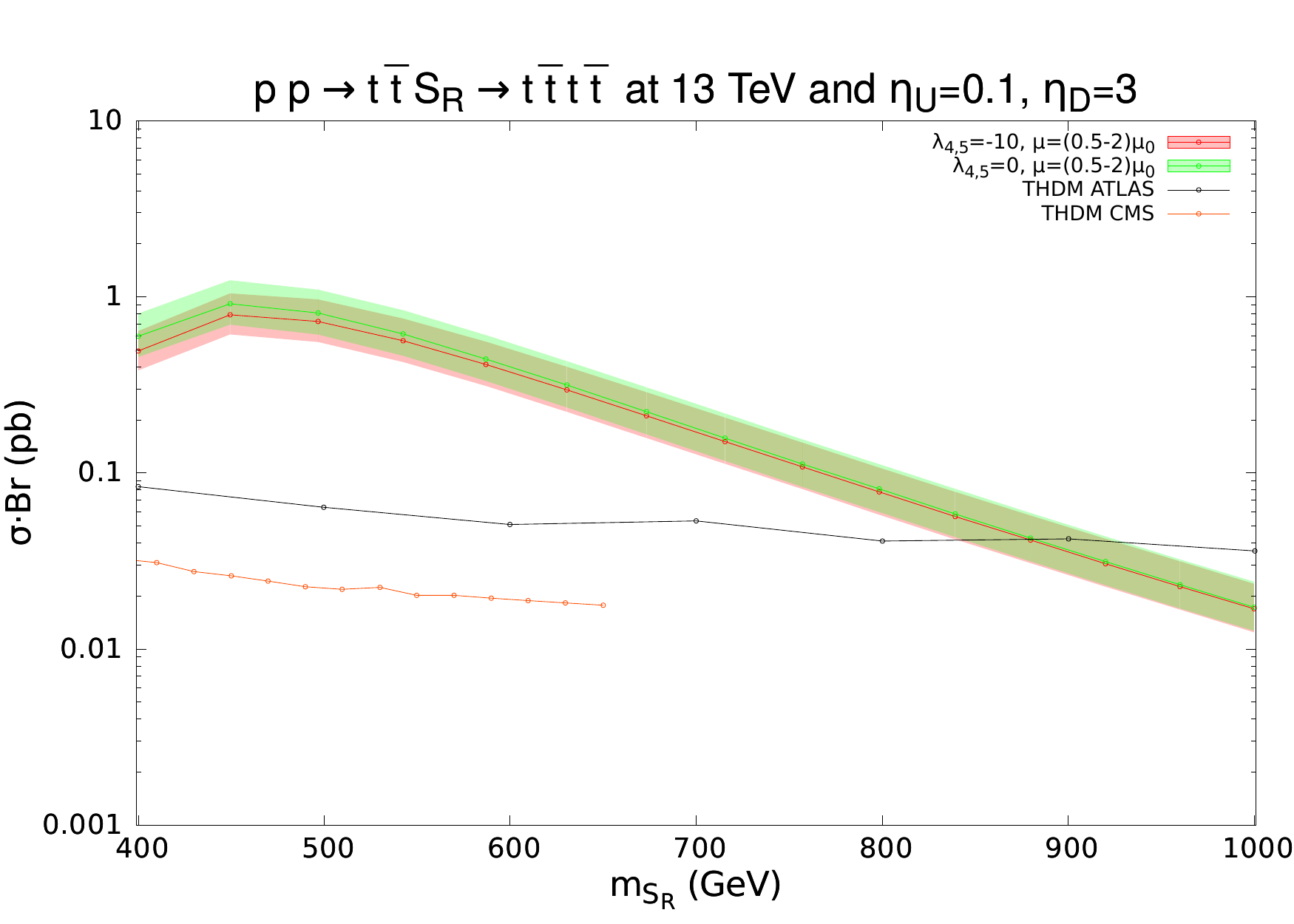}
}\\
\subfloat[]{
\includegraphics[scale=0.5]{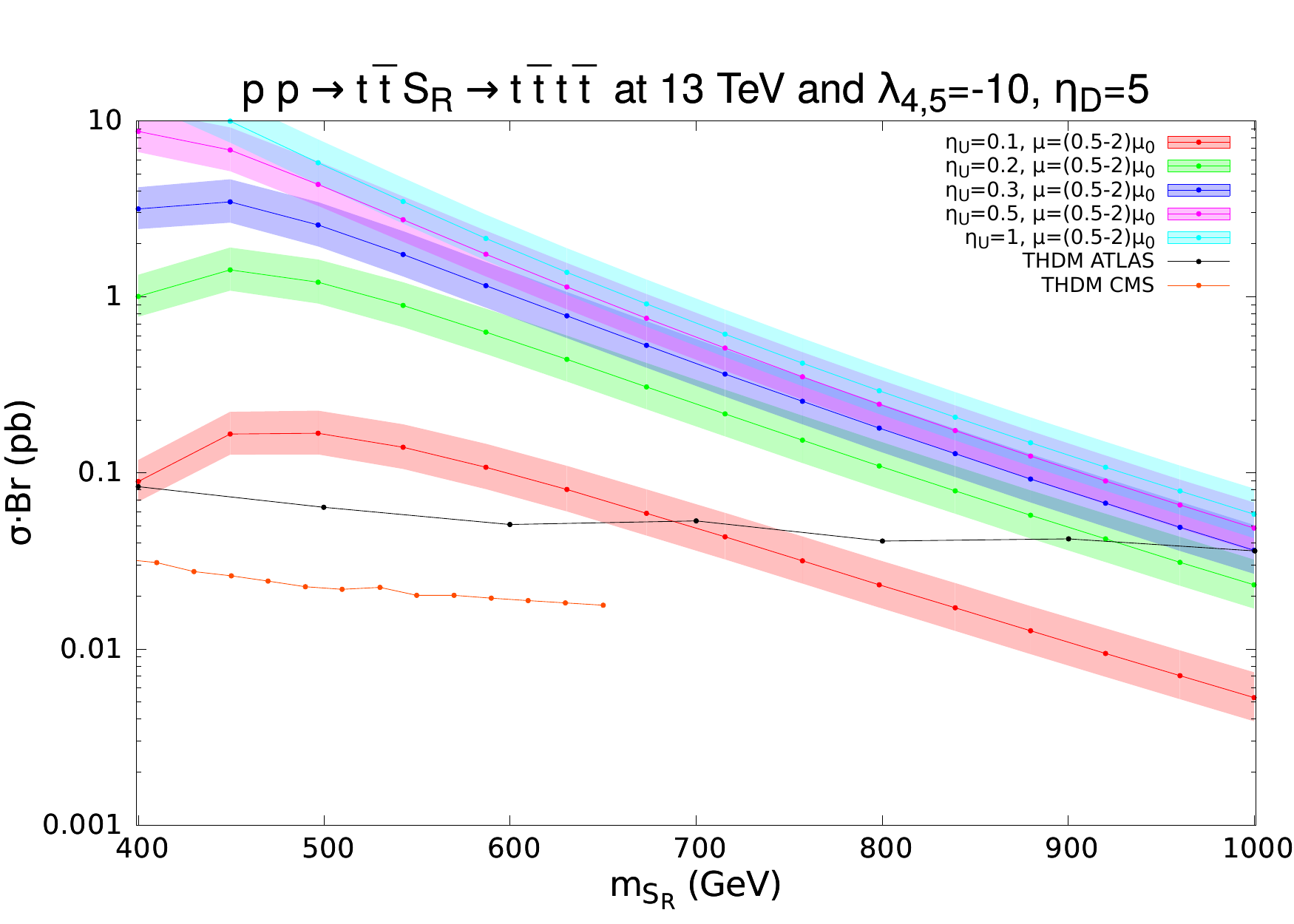}
}
\subfloat[]{
\includegraphics[scale=0.5]{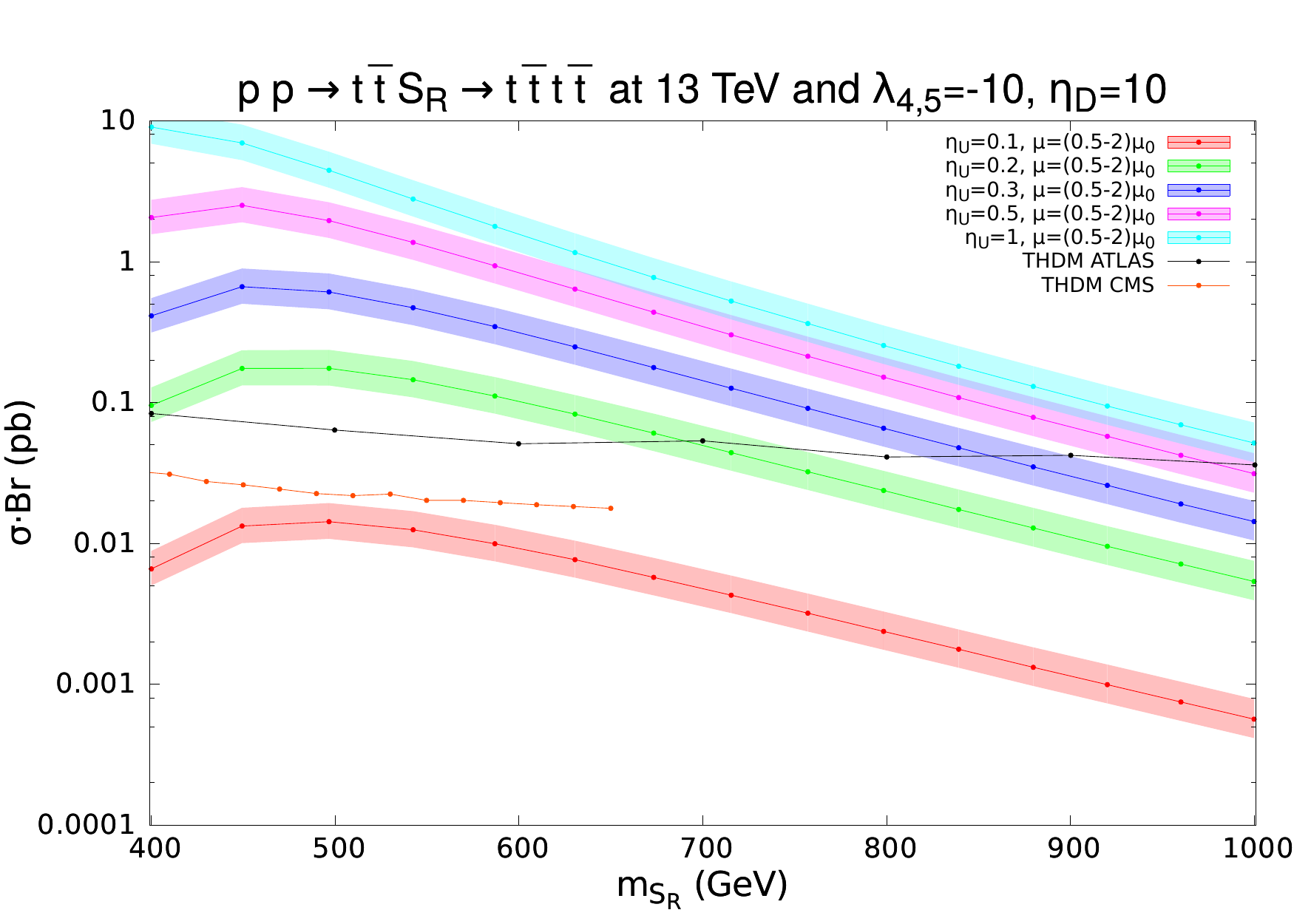}
}
\caption{Cross section of the associated production of $S^0_R$ with $t \bar{t}$ times its branching ratio into $t \bar{t}$, as a function of $m_{S^0_R}$, for representative choices of the parameters. In the top panels $\eta_U=0.1$ and $\lambda_{4,5}$ is varied from 0 to $-10$, while
$\eta_D=1$ (left) or $\eta_D=3$ (right). The bottom panels correspond to $\lambda_{4,5}=-10$, with 
$\eta_U$ varying from 0.1 to 1, and $\eta_D=5$ (left) or $\eta_D=10$ (right). The experimental bounds are taken from Refs.~\cite{Aaboud:2018xpj} and \cite{Sirunyan:2019wxt}.
}
\label{fig:Stt4t} 
\end{figure}
%%%%%%%%%%%%%%%%%%%%%%%%%%%%%%%%%%%%%%%%%%%%%%%%%%%%%%%%%%%%%%%%%%%%%%%

%%%%%%%%%%%%%%%%%%%%%%%%%%%%%% Figure %%%%%%%%%%%%%%%%%%%%%%%%%%%%%%
 \begin{figure}[t!]
\centering
\subfloat[]{
\includegraphics[scale=0.5]{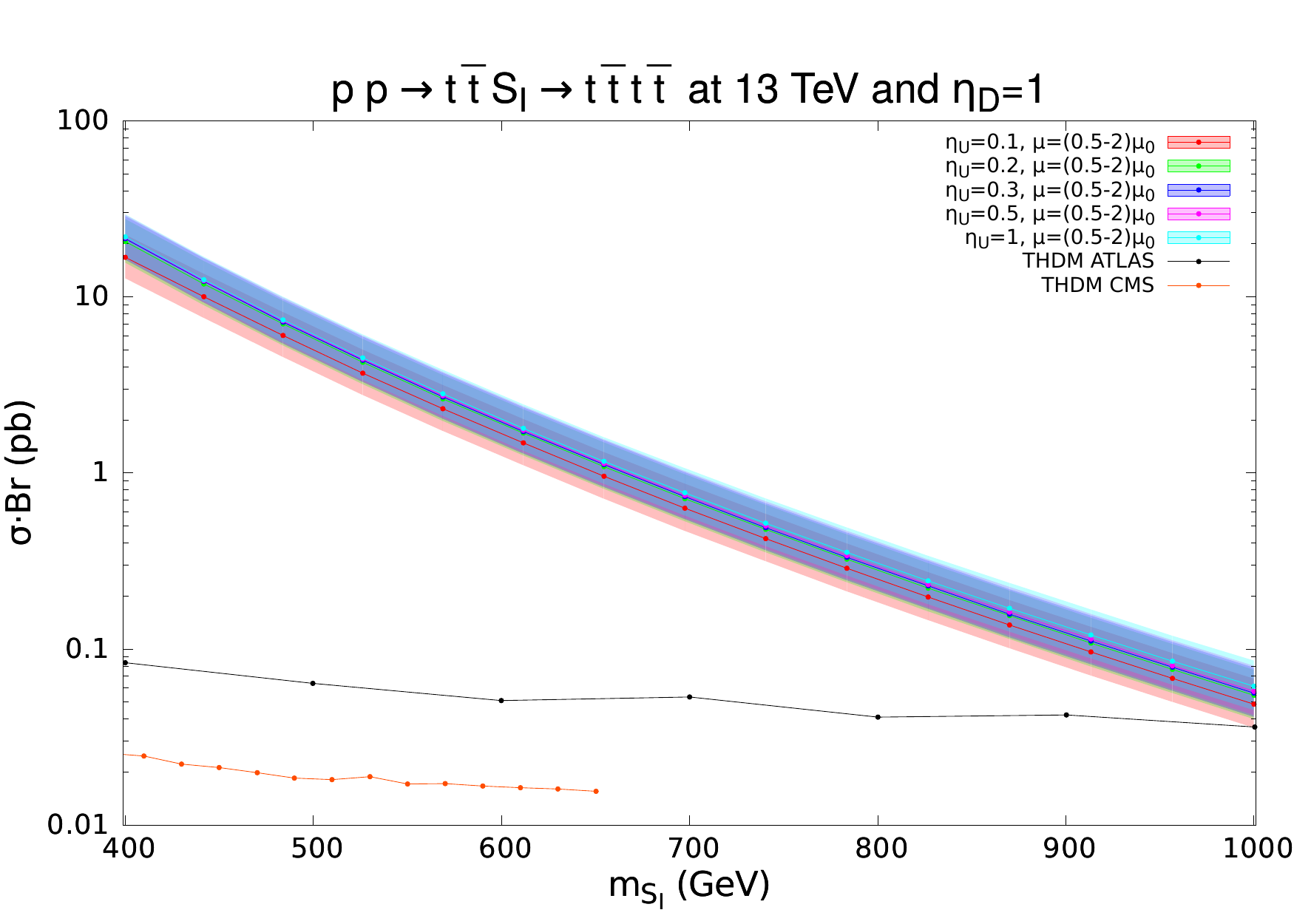}
}
\subfloat[]{
\includegraphics[scale=0.5]{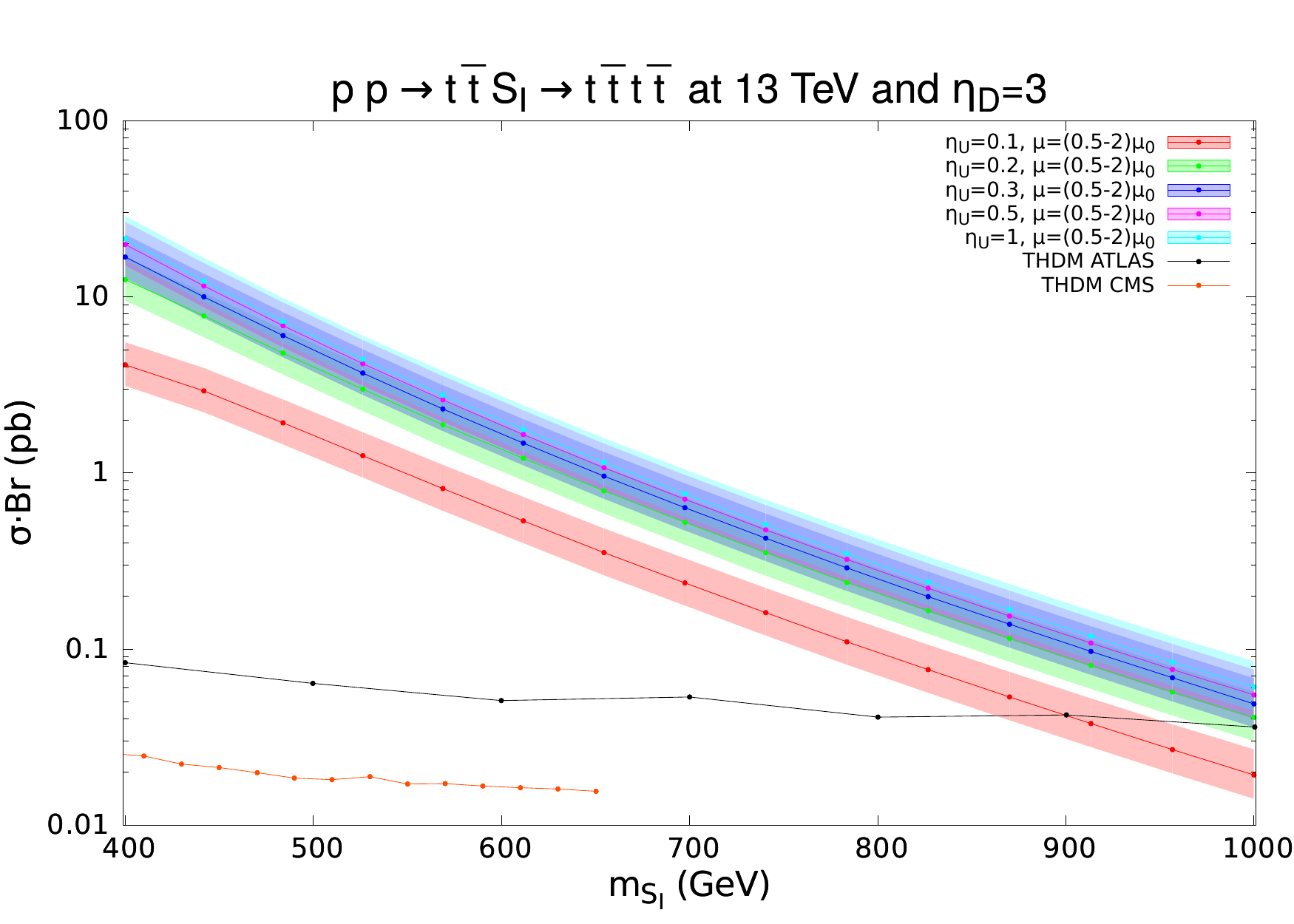}
}\\
\subfloat[]{
\includegraphics[scale=0.5]{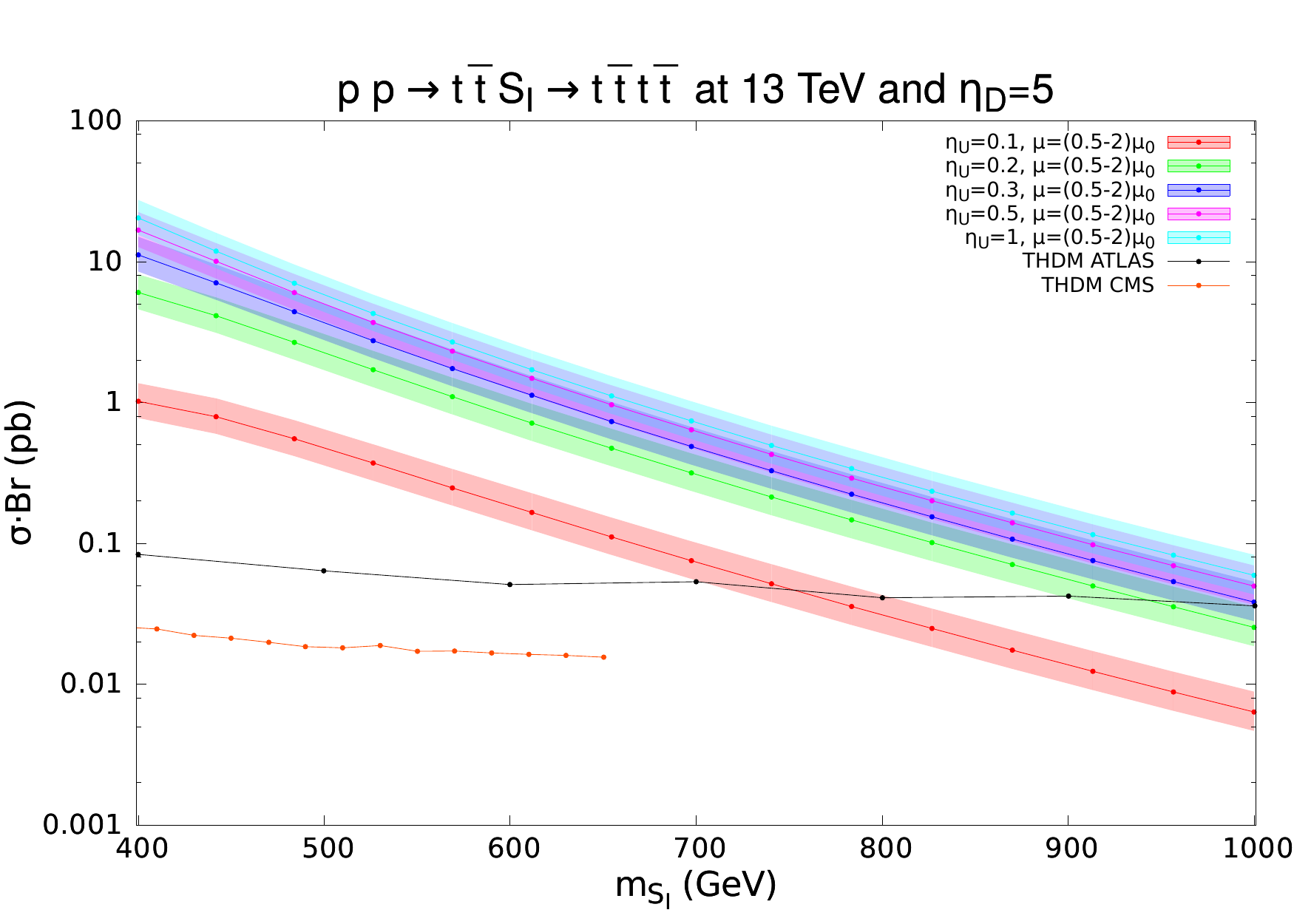}
}
\subfloat[]{
\includegraphics[scale=0.5]{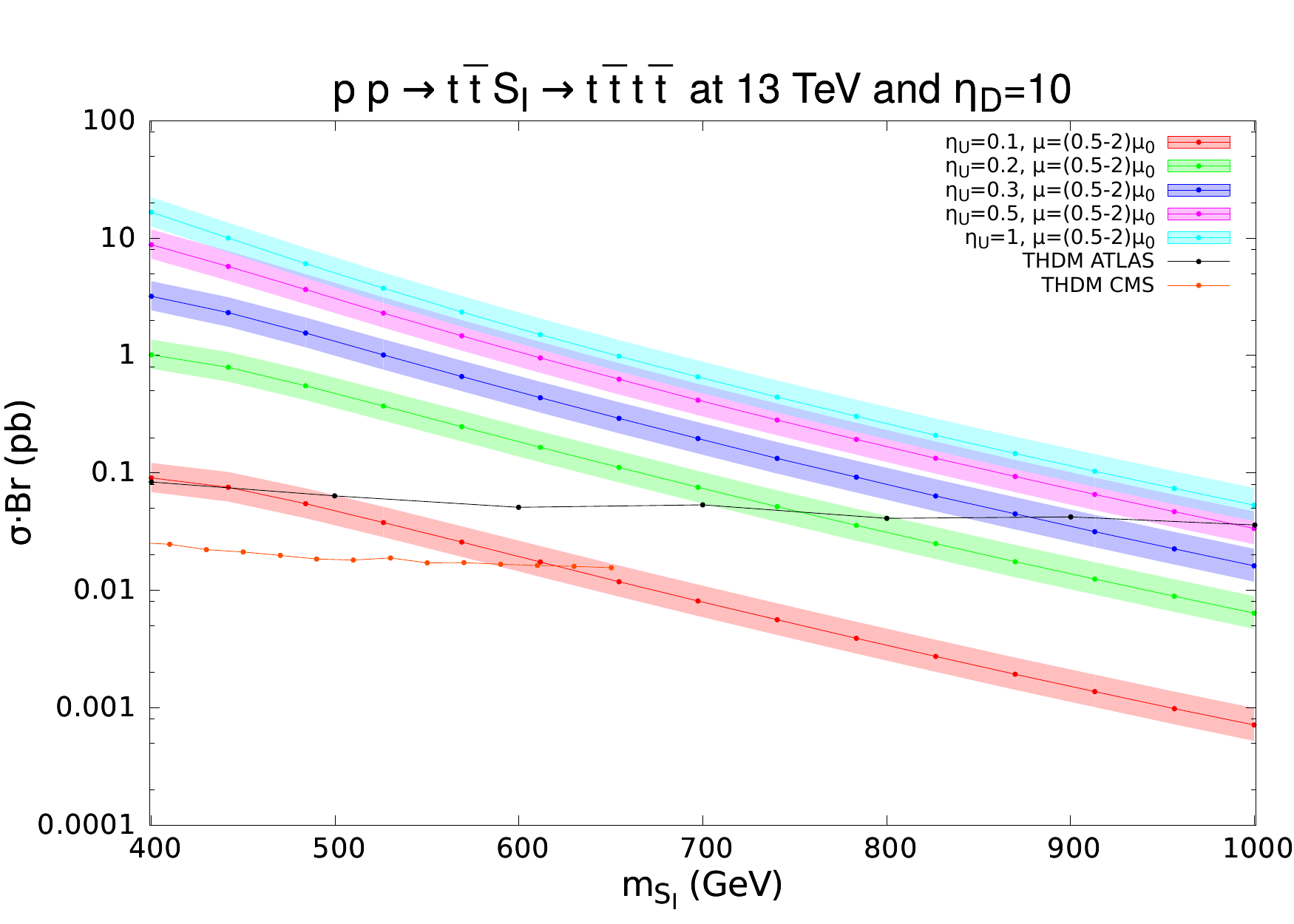}
}
\caption{Cross section of the associated production of $S^0_I$ with $t \bar{t}$ times its branching ratio into $t \bar{t}$, as a function of $m_{S^0_I}$, for representative choices of the parameters. In all panels, $\eta_U$ is varied from $0.1$ to 1, while
$\eta_D=1$ (top-left), 3 (top-right), 5 (bottom-left) and 10 (bottom-right). The experimental bounds are taken from Refs.~\cite{Aaboud:2018xpj} and \cite{Sirunyan:2019wxt}.
}
\label{fig:Sitt4t} 
\end{figure}
%%%%%%%%%%%%%%%%%%%%%%%%%%%%%%%%%%%%%%%%%%%%%%%%%%%%%%%%%%%%%%%%%%%%%%%

The production of the coloured neutral scalars in association with top quarks proceeds through the two mechanisms indicated in Fig.~\ref{fig:Diagrams_Stt}. Both diagrams contain a single $S^0_{R,I} t\bar t$ vertex with a coupling $\eta_U$. However, in the kinematical region of interest, the production through the left mechanism is dominated by an on shell intermediate $S^0_{R,I}$ particle decaying into $t\bar t$.
Therefore, as long as the branching ratio of this decay is close to one, the dependence with $\eta_U$ will be small (gets cancelled by the total decay width contained in the on shell propagator). The second mechanism on the right will then become more relevant for values of $|\eta_U|$ in its higher allowed range. However, for values of $|\eta_U|$ of order $10^{-1}$ or smaller, the left diagram dominates, provided the branching ratio of the decay to top quarks is of order one. Note that when the left diagram dominates, the resulting amplitude does not change under a global rescaling of the parameters $\eta_U$, $\eta_D$ and $\lambda_{4,5}$.\footnote{Obviously, the decay width into top quarks should be large enough for the decay to happen before the detector. At the LHC, this condition is fulfilled even for values of $|\eta_U|$ of order $10^{-7}$.}

The limits on the production of neutral scalars in association with top-quark pairs are shown in Figs.~\ref{fig:Stt4t} and  \ref{fig:Sitt4t}, for the CP-even and CP-odd cases, respectively. As mentioned before, we compare our theoretical predictions
for $\sigma\cdot\mathrm{Br}$ of \ $pp\to S^0_{R,I}\, t\bar t\to t\bar t\, t\bar t$ \
with the experimental upper bounds obtained for the type-II 2HDM by ATLAS \cite{Aaboud:2018xpj} and CMS \cite{Sirunyan:2019wxt}.\footnote{Diagram \ref{fig:Diagrams_Stt}a is absent in the 2HDM. Since this topology dominates at small values of $|\eta_U|$, we have simulated the kinematical cuts employed by ATLAS~\cite{Aaboud:2018xpj} and checked that the selection efficiency is not lower in our case.}
Note that the ATLAS analysis covers a higher mass region, reaching masses up to 1 TeV, while the CMS one, although being more restrictive, only applies to $m_{S^0_{R,I}}\le 650$~GeV. Regarding the QCD corrections, they increase the SM prediction for this process ($pp\to H\, t\bar t\to t\bar t\, t\bar t$) by a factor of 1.2 \cite{Beenakker:2002nc}. We expect them to be more relevant for the coloured scalars, because they can directly couple to gluons. However, these corrections have not been calculated yet. Since they are expected to increase the prediction for $\sigma\cdot\mathrm{Br}$, our limits without QCD corrections are then conservative.

The four panels in Fig.~\ref{fig:Stt4t} show the predicted production of the CP-even scalar, as a function of $m_{S^0_R}$, for several representative choices of the relevant parameters. 
In all cases, we have considered that the CP-even neutral scalar is the lightest coloured scalar.
Taking 
$\eta_U = 0.1$ and $\eta_D\leq1$ (top-left panel), the predictions are well above the experimental bounds in the whole range of $m_{S^0_R}$, for any value of $|\lambda_{4,5}|$, resulting in a lower limit of 1 TeV for the mass of the neutral scalar. This limit remains valid for larger values of $|\eta_U|$ because they result in much larger theoretical predictions for $\sigma\cdot\mathrm{Br}$.
The expected signal decreases with increasing values of $|\eta_D|$, mainly because it increases the branching ratio of the decay to bottom quarks, relaxing the experimental constraint. The  decay amplitude to gluons also depends on $\eta_D$, and on its relative sign with $\eta_U$ and $\lambda_{4,5}$, but the effect is extremely weak. Indeed, the decay to gluons is also affected by $\lambda_{4,5}$ but, as we discussed before, the effect is small for values of this parameter inside its perturbative unitarity region. Looking at the top panels, we can see that the difference between setting $\lambda_{4,5}$ to 0 or to -10 is very small.  For positive values of $\lambda_{4,5}$, the effect would be even smaller because the decay amplitude to gluons would grow less while varying $\lambda_{4,5}$. With $\eta_U = 0.1$ and $\eta_D=3$ (top-right panel), one still obtains $m_{S^0_R}\ge 850$~GeV, for values of $\lambda_{4,5}$ inside its perturbative unitarity region. As we keep increasing $\eta_D$ (lower panels) the constraints become worse, but even with $\eta_D= 10$ the mass of the CP-even scalar is constrained to be higher than 1~TeV for $\eta_U=1$.

For the analysis of the CP-odd neutral scalar, we have chosen $\lambda_2$ and $\lambda_3$ in such a way that this particle is the lightest coloured scalar, as mentioned before. As illustrated by Fig.~\ref{fig:Sitt4t}, the results are similar to the previous ones; although in this case, there is no dependence on $\lambda_{4,5}$, which makes the analysis simpler.  The decay width into gluons can be neglected compared to the decays into top and bottom quarks because it is a loop process with only top and bottom quarks in the loop. 
The dependence of the resulting limits with $\eta_U$ and $\eta_D$ follows the same trend as before: the higher the value of $|\eta_U/\eta_D|$, the better the constraint. We find again a lower limit of 1 TeV for values of $|\eta_U|$ as small as $10^{-1}$ when $\eta_D=1$ (top-left panel). For $\eta_U = 0.1$ and $\eta_D=3$ (top-right panel), the constraint is $m_{S^0_I}\ge 850$~GeV, as in the CP-even case. However, thanks to the CMS data, we can find a better limit
on the CP-odd scalar, $m_{S^0_I}\ge 575$~GeV, for $\eta_D=10$ and $\eta_U=0.1$ (bottom-right panel).

\subsection{Production of Charged Scalars}

The analysis of charged-scalar production is much simpler because they cannot decay to gluons; therefore, there is no dependence on $\lambda_{4,5}$.
Now, we restrict ourselves to the region of parameter space with $\lambda_2 + 2\lambda_3 > 0$ and $\lambda_2 - 2\lambda_3 > 0$, so that the decays  $ S^+ \rightarrow S_{R,I}^0 W^+ $ are kinematically forbidden. Thus, the only possible two-body decays of the charged scalars are into two quarks, and therefore, the decay to heavy quarks will always dominate. Hence, the cross section only depends on $\eta_U$ and $\eta_D$.

%%%%%%%%%%%%%%%%%%%%%%%%%%%%%% Figure %%%%%%%%%%%%%%%%%%%%%%%%%%%%%%%
\begin{figure}[t]
\centering
\subfloat[]{
\includegraphics[scale=0.28]{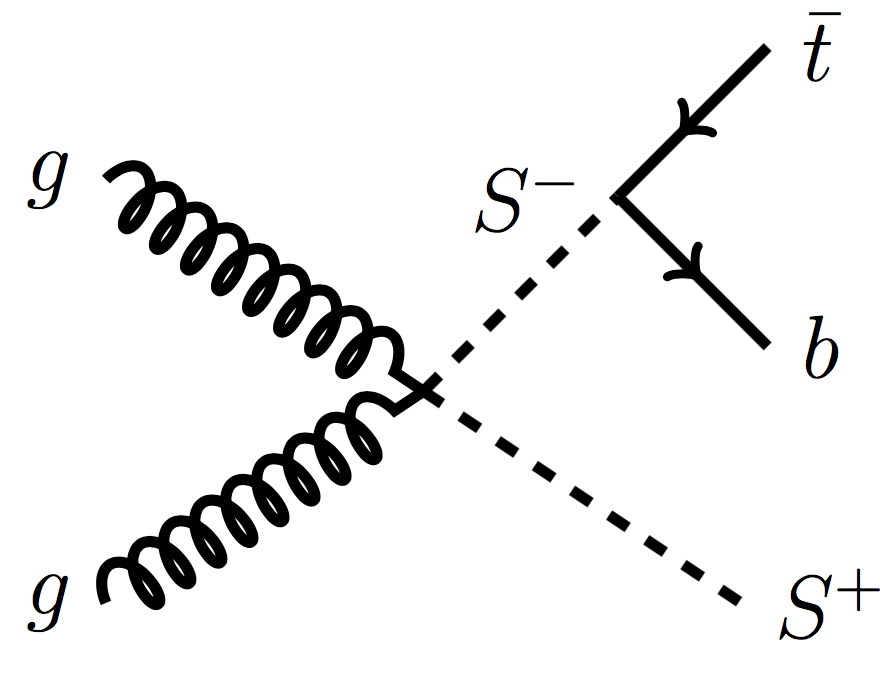}
}
\hspace{2.0 cm}
\subfloat[]{
\includegraphics[scale=0.28]{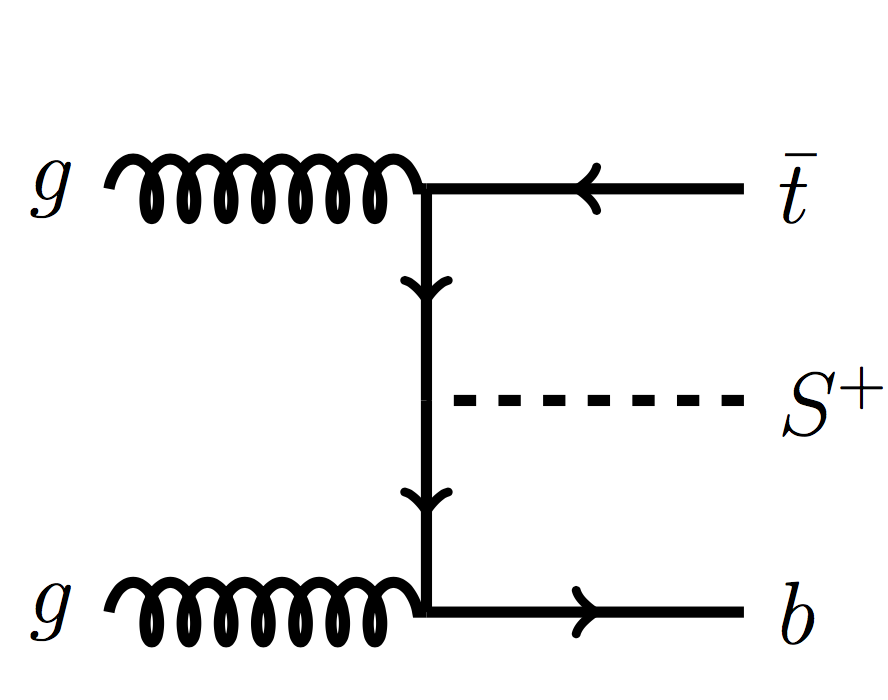}
}
\caption{Representative Feynman diagrams contributing to the associated production of charged scalars with heavy quarks.}
\label{fig:Diagrams_Stb} 
\end{figure}
%%%%%%%%%%%%%%%%%%%%%%%%%%%%%%%%%%%%%%%%%%%%%%%%%%%%%%%%%%%%%%%%%

The production mechanisms, indicated in Fig.~\ref{fig:Diagrams_Stb}, are analogous to the 
associated production of a neutral scalar with the obvious changes on the particle charges. Again, the mechanism on the right dominates when $|\eta_U|\sim\mathcal{O}(1)$, while the left diagram is the relevant one for $|\eta_U|$ of order $10^{-1}$ or smaller. But now the branching ratio of the decay into top and bottom quarks is always close to one, independently of the values of $\eta_U$ and $\eta_D$, provided the charged scalar is not fermiophobic ($\eta_{U,D}=0$). The only requirement needed for our analysis is that the charged scalar indeed decays before entering the detector,
which is the case even when $\eta_U$ and $\eta_D$ are of order $10^{-5}$.

%%%%%%%%%%%%%%%%%%%%%%%%%%%%%% Figure %%%%%%%%%%%%%%%%%%%%%%%%%%%%%%%
\begin{figure}[tb]
\centering
\subfloat[]{
\includegraphics[scale=0.5]{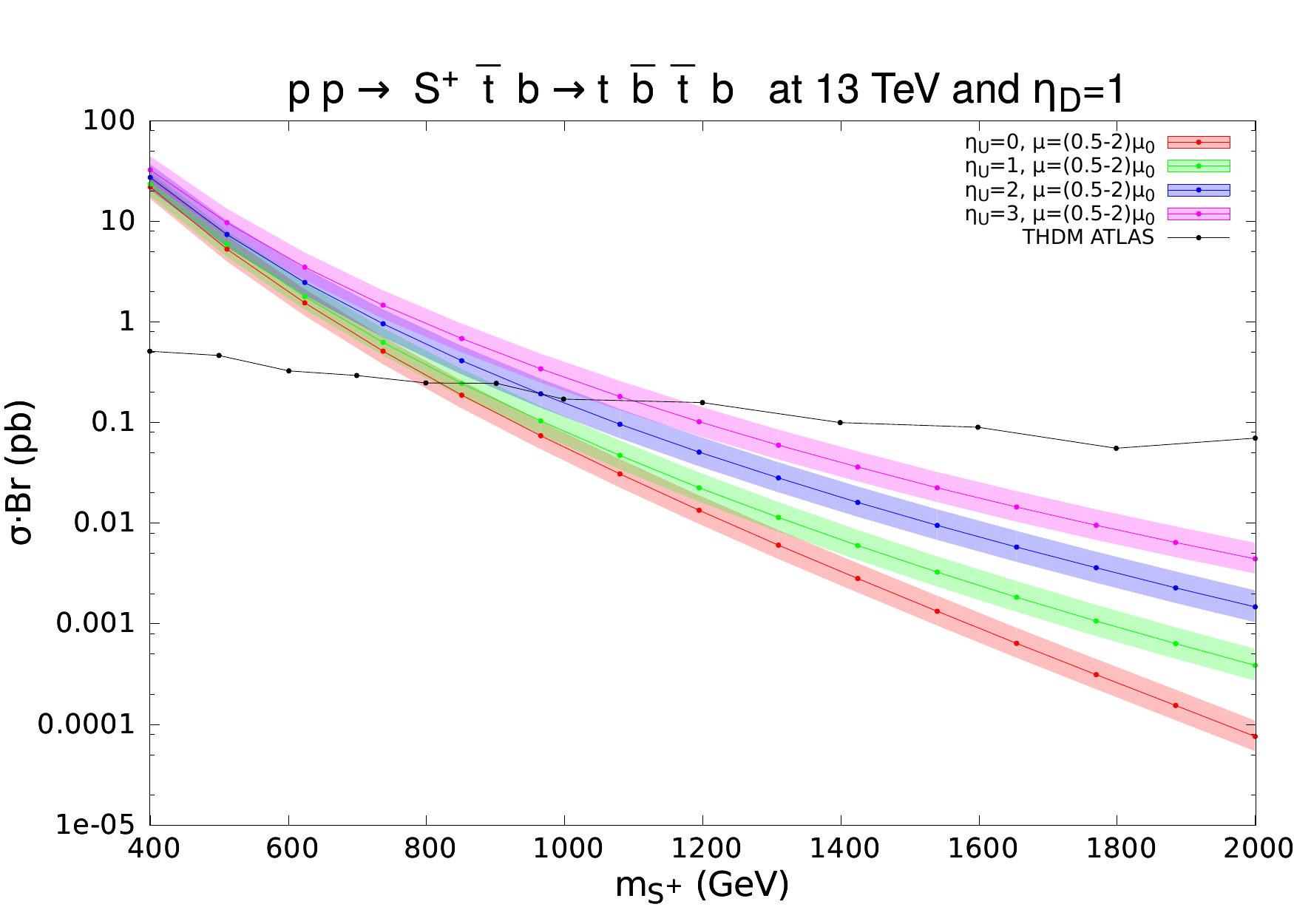}
}
\subfloat[]{
\includegraphics[scale=0.5]{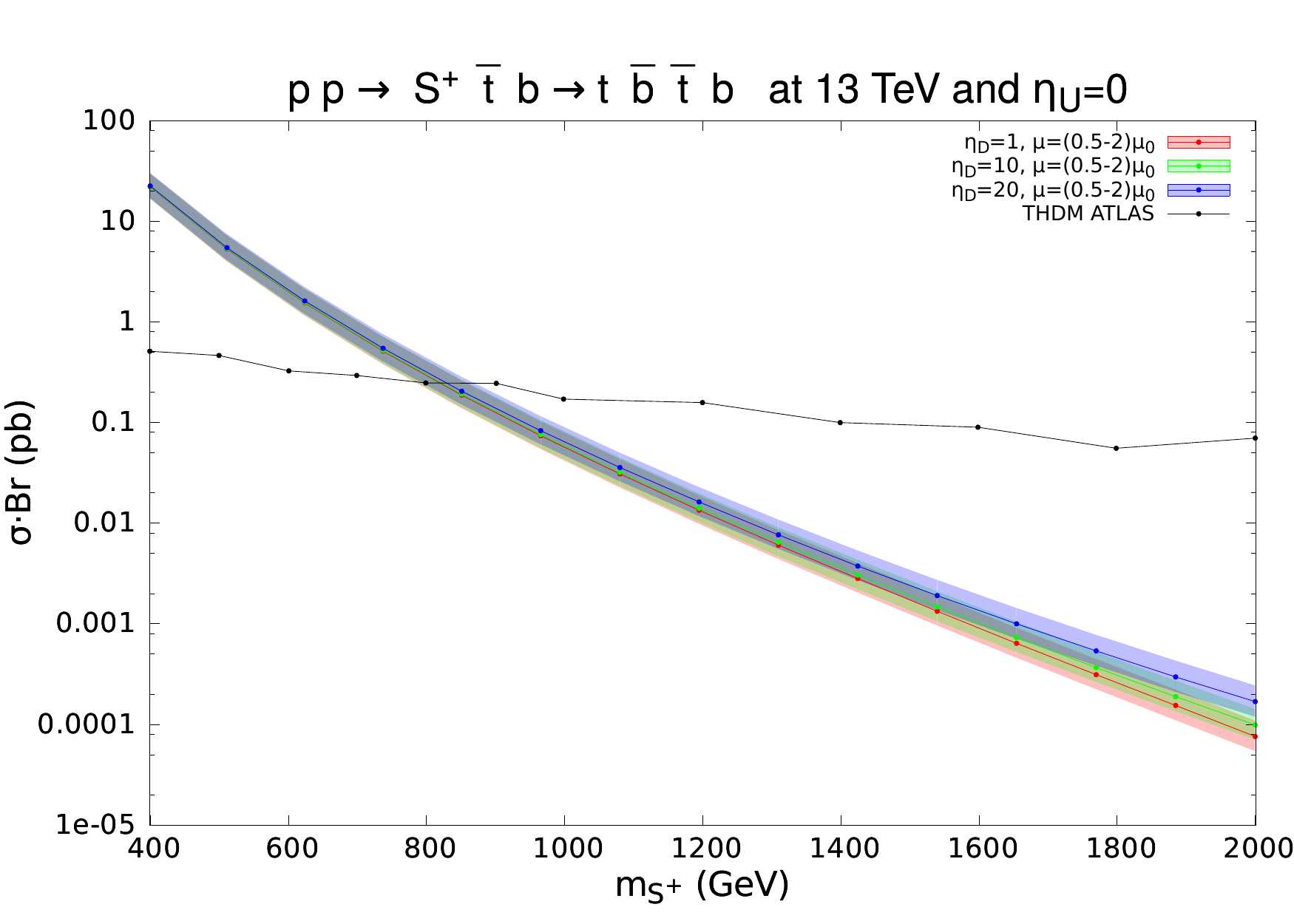}
}
\caption{
$\sigma\cdot\mathrm{Br}$ of the process $pp\to S^+\bar t\, b\to t\,\bar b\: \bar t\, b$, as a function of $m_{S^\pm}$, compared with the (95\% C.L.) experimental bound~\cite{Aaboud:2018cwk}.}
\label{fig:Stb} 
\end{figure}
%%%%%%%%%%%%%%%%%%%%%%%%%%%%%%%%%%%%%%%%%%%%%%%%%%%%%%%%%%%%%%%%%%%%%%%%%

Fig.~\ref{fig:Stb} compares our theoretical predictions for the cross section times branching ratio of the process $pp\to S^+\bar t\, b\to t\,\bar b\; \bar t\, b$, with the (95\% C.L.) experimental bounds, as a function of $m_{S^\pm}$. We profit here from the ATLAS analysis performed in the context of the type-II 2HDM model~\cite{Aaboud:2018cwk}. In the left panel, $\eta_D=1$ and $\eta_U$ is varied from 0 to 3, while in the right panel the most pessimistic possibility, $\eta_U=0$, is taken and $\eta_D$ is varied between 1 and 20.
From this plot, we are able to constrain the mass of the charged scalars to be higher than 800 GeV, even for $\eta_U=0$ and for any value of $|\eta_D|$ as small as $10^{-5}$. For higher values of $|\eta_U|$ the constraints become stronger, and for $\eta_U=2$  we push this limit up to 900 GeV. Note that this is the maximum value of $\eta_U$ allowed by flavour constraints, for
charged-scalar masses up to 1 TeV  \cite{gw07,clyz15}, with a 95\% C.L. 

The production cross section depends very weakly on $|\eta_D|$. This is easy to understand, since $\eta_D$ couples proportionally to the bottom quark mass while $\eta_U$ brings an $m_t$ factor. Even when $\eta_U=0$ (right panel), the sensitivity to  $|\eta_D|$ is quite mild because 
the left diagram dominates in this case.

\section{Conclusion}
\label{sec:Summary}

In this work, we have found lower limits for the masses of all the coloured scalars of the MW model: the CP-even and CP-odd neutrals and the charged coloured scalars. Our phenomenological study has been performed in the CP-conserving limit. In order to avoid unwanted decay modes into other coloured scalars, for each separate analysis, we have selected the parameters of the scalar potential in such a way that the analyzed scalar was the lightest. Thus, once we find a limit for the mass of this scalar, the others must be necessarily heavier, so the limit applies to all of them. Combining the results obtained for all possible mass splittings, the  least restrictive limit is valid for any value of the parameters $\lambda_2$ and $\lambda_3$.

We have analyzed the single production of neutral scalars and the associated production of neutral and charged scalars with heavy quarks, using the available LHC data at $\sqrt{s}=13$~TeV.
From the study of the single production, we could not find better constraints than those already obtained in Ref.~\cite{Hayreter:2017wra}, at lower LHC energies. However, our analysis of the associated production results in relevant limits on the scalar masses, which are significantly better than those extracted in previous works.

The associated production of the CP-even neutral scalar depends on $\lambda_{4,5}$, $\eta_U$ and $\eta_D$. Nevertheless, we have found a very small sensitivity to $\lambda_{4,5}$ for values of $|\eta_U|$ higher than $10^{-1}$. If this is the case, the limits on the scalar mass only depend on the relation between $\eta_U$ and $\eta_D$. For $|\eta_D/\eta_U| \le 10$, a CP-even scalar is excluded in the whole range explored by ATLAS, which extends up to 1 TeV. The limit gets relaxed for larger values of this ratio. Taking  $|\eta_D/\eta_U| \sim 30$, one still finds $m_{S^0_R} > 850$~GeV. However, no limit is found when $|\eta_D/\eta_U| > 100$. Similar results are obtained for
$|\eta_U|$ smaller than $10^{-1}$, as long as $|\lambda_{4,5}/\eta_U| \le 100$.

The associated production of the CP-odd scalar behaves in a similar way, but it does not depend on $\lambda_{4,5}$, which results in somewhat stronger constraints for some parameter configurations. The limits on $m_{S^0_I}$ are only sensitive to the ratio $\eta_D/\eta_U$ and are valid for values of $\eta_U$ as small as $10^{-7}$. For $|\eta_D/\eta_U| \le 10$, one finds again that a CP-odd scalar is excluded in the full kinematical range analyzed, i.e., $m_{S^0_I} > 1$~TeV.
The limit gets relaxed to 850 GeV, when $|\eta_D/\eta_U| \sim 30$, as in the CP-even case. However, even for $|\eta_D/\eta_U| \sim 100$, one still finds a relevant limit of $m_{S^0_I} > 575$~GeV.

Since the charged scalars cannot decay into gluons, the analysis of their associated production is much simpler. We found a quite strong lower bound, $m_{S^\pm} > 800$~GeV, provided the charged scalars are not fermiophobic. This constraint remains valid as long as  $|\eta_D| > 10^{-5}$ or $|\eta_U|>10^{-7}$. The limit becomes stronger with increasing values of $|\eta_U|$, reaching 900~GeV for $|\eta_U|=2$, the maximum value of this parameter allowed by flavour constraints.

Combining all searches, we find an absolute lower bound of 800~GeV on the masses of all coloured scalars, which is valid under very mild requirements on the relevant model parameters: 
$|\eta_D/\eta_U| < 30$, $|\lambda_{4,5}/\eta_U| < 100$ and $|\eta_U|>10^{-7}$. Note that this limit
applies for all possible mass splittings among the scalars. Our analysis puts then a very severe limitation on the hypothetical existence of light coloured scalars with masses below the TeV. In practice, they seem to be strongly excluded for all reasonable choices of parameters. The only possibility that remains still viable are fermiophobic scalars with $\eta_{U,D}=0$. The analysis of this extreme case would require a quite different and much more tricky phenomenological approach. 

Much larger data samples are going to be accumulated in the forthcoming run 3 and, specially, at the High Luminosity LHC (HL-LHC). Prospective analyses for the pair production of pseudoscalar colour-octet particles decaying into the $(t\bar t) (t\bar t)$ final state \cite{Darme:2018dvz} have been presented in Ref.~ \cite{Azzi:2019yne}. From these results, we can infer that our 850 GeV limit on the CP-odd scalar mass, when $|\eta_D/\eta_U| \sim 30$, could be pushed to 1~TeV. In order to find a rough estimate for the High Energy LHC (HE-LHC), we have generated some events at a center-of-mass energy of 27 TeV in the same channel, finding an exclusion lower limit of 1.4 TeV in the top-philic case. Larger mass scales could of course be reachable at the FCC-hh, since the pair production cross section would be orders of magnitude larger, for a given scalar mass value \cite{Abada:2019lih}. In the single production channel, the achievable sensitivity at the FCC-hh can be roughly estimated from existing simulations of strongly interacting vector particles, which quote limits on
$\sigma(pp\to Z')\times \mathrm{Br}(Z'\to t\bar t)$ in the $10^{-4}$ to $10^{-3}$~pb range, for $Z'$ masses between 10 and 35 TeV \cite{Abada:2019lih}. Extrapolating these estimates to lower masses, exclusion limits around 3 to 5 TeV for the octet scalar mass appear to be feasible.
More detailed analyses would be needed to reliably assess the discovery potential of future colliders.

\section*{Acknowledgements}
We would like to thank A. Pe\~nuelas for her useful comments on the manuscript. We also thank M. Perello and M. Vos for helpful discussions on experimental aspects of the analysis.
This work has been supported in part by the Spanish Government and ERDF funds from
the EU Commission [grant FPA2017-84445-P], the Generalitat Valenciana [grant Prometeo/2017/053] and the Spanish Centro de Excelencia Severo Ochoa Programme [grant SEV-2014-0398].
The work of V.M. is supported by the FPU doctoral contract FPU16/0191, funded by the Spanish Ministry of Science, Innovation and Universities.

\bibliography{LHC_bounds_on_coloured_scalars}
\bibliographystyle{utphys}

\end{document}